\documentclass[preprint]{elsarticle}
\usepackage{amsmath,amssymb,amsfonts}
\usepackage{natbib}
\usepackage{algorithmic}
\usepackage{graphicx}
\usepackage{textcomp}
\usepackage{enumitem}
\usepackage{caption}
\usepackage{hyperref}

\journal{NIM-A}

\date{\today}
\begin{document}

\begin{frontmatter}

\title{Position-Sensitive Silicon Photomultiplier Array with Enhanced Position Reconstruction by means of a Deep Neural Network}

\author[1]{Cyril Alispach\corref{cor1}}
\ead{cyril.alispach@unige.ch}
\author[4]{Fabio Acerbi}
\author[3]{Hossein Arabi}
\author[1]{Domenico della Volpe}
\author[4]{Alberto Gola}
\author[1]{Aramis Raiola}
\author[3,5,6,7]{Habib Zaidi}

\affiliation[1]{
    organization={Départment de Physique Nucléaire et Corpusculaire, Université de Genève}, 
    adressline={quai Ernest-Ansermet 24},
    postcode={1211}, 
    city={Genève 4}, 
    country={Switzerland}}

\affiliation[3]{
    organization={Division of Nuclear Medicine and Molecular Imaging, Geneva University Hospital},
    postcode={1205},
    city={Geneva}, 
    country={Switzerland}}

\affiliation[4]{
    organization={Center for Sensor and Devices (SD), Fondazione Bruno Kessler},
    adressline={Via Sommarive 18},
    postcode={38123},
    city={Trento}, 
    country={Italy}}

\affiliation[6]{
    organization={Department of Nuclear Medicine and Molecular Imaging, University of Groningen, University Medical Center Groningen},
    postcode={9700 RB},
    city={Groningen}, 
    country={Denmark}}
\affiliation[7]{
    organization={Department of Nuclear Medicine, University of Southern Denmark},
    postcode={DK-500},
    city={Odense}, 
    country={Denmark}}
    
\affiliation[5]{
    organization={University Research and Innovation Center, Obuda University},
    city={Budapest}, 
    country={Hungary}}
    
\cortext[cor1]{Corresponding author}

\begin{abstract}

Single-photon sensitive detectors like Silicon Photomultipliers are widely used in many medical imaging applications. By using detectors with position resolutions, it is possible to build compact photodetector readouts with reduced number of channels, but still preserving position resolution and gamma-rays imaging capabilities. In this work, we present the advantage of using a Deep Neural Networks (DNNs) light position reconstruction applied to a 2x2 array of linearly-graded SiPMs (LG-SiPMs), to minimize the distortions on the reconstructed event maps.
Our approach significantly enhances both the resolution and linearity of position detection compared to the nominal reconstruction formula based on the device architecture.
Remarkably, the DNN-based reconstruction boosts the number of resolved areas (‘pixels’) by a factor of 5.7 to 12.1 (depending the training splitting used) allowing for a higher level of precision and performance in light detection.

\end{abstract}




\end{frontmatter}

\section{Introduction}

Gamma cameras are widely used in medical imaging and other applications, typically based on high-sensitivity photo-detectors that read out scintillator crystals. Examples are Single photon emission computed tomography (SPECT) and positron emission tomography (PET)~\cite{AngerCamera, gamma-camera-performance, Peterson2011, Tsuchimochi_2013, Farnworth2023}. In such applications, position reconstruction is an important part of the acquisition and affect the performance of the overall imaging technique. The scintillating crystal converts gamma-rays into a detectable number of optical photons, which are then converted into an electric signal by the photodetector. The new generation of gamma-ray imaging devices are often employing silicon photomultipliers (SiPMs) as they offer high internal gain, good photo-detection efficiency, good timing resolution, and robustness compared to other detectors like PMTs~\cite{Farnworth2023, MedicalSiPM, Gundacker_2020, Acerbi2025, Della_Volpe2025-qd}. 

\begin{figure}
    \centering
    \includegraphics[width=0.45\linewidth]{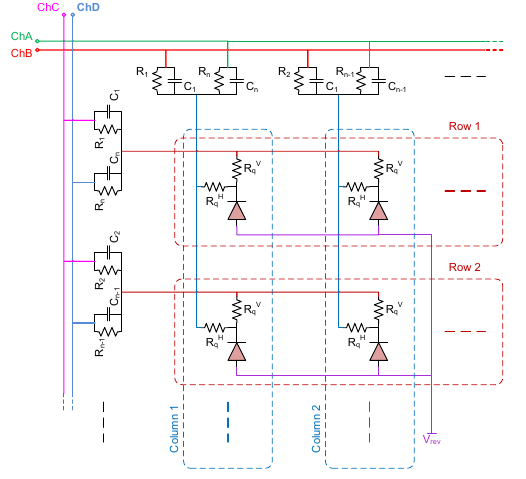}
    \includegraphics[width=0.5\linewidth]{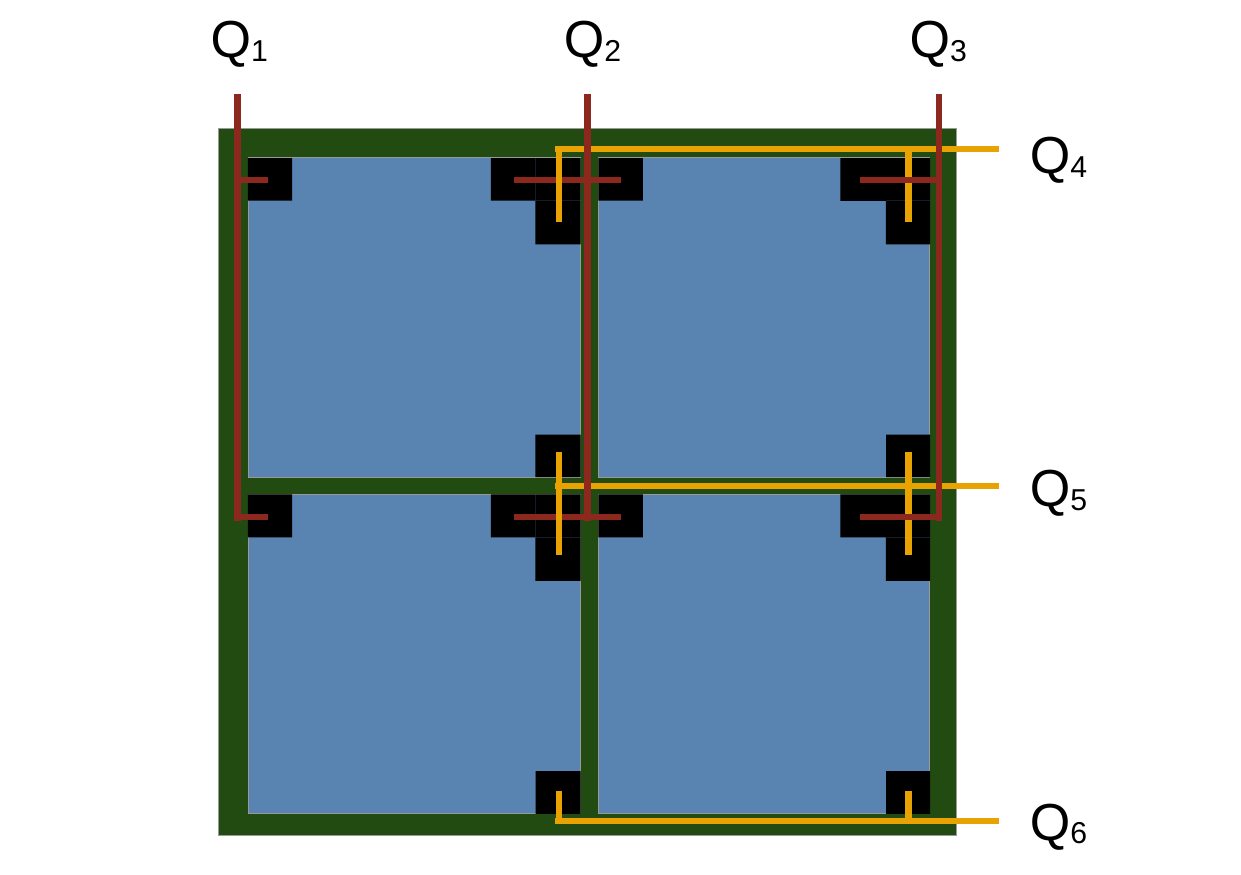}
    \caption{Left: Scheme showing a portion of a single LG-SiPM chip's micro-cells connected to the resistive/capacitive network allowing to reconstruct the hit position from 4 channels. Taken from \cite{Acerbi2024}. Right: Connection scheme of the 2x2 LG-SiPM elements arranged in a "smart channel" approach. The position can be reconstructed with equation \ref{eq:relative_coordinates}. Adapted from \cite{Acerbi2024}.}
    \label{fig:schematic_channels}
\end{figure}

Usually, position reconstruction is obtained by having multiple detectors arranged in a 2D resistive network, or using a multi-anode PMT (MA-PMT) with the multi-anode readout with resistive network~\cite{Zhang2023, Schaefer1998} such that the position can be inferred from the relative amplitudes of the signals in only 4 channels. This charge-based Center of Gravity (CoG) approach was pioneered in the 50s by Hal O. Anger~\cite{AngerCamera}. This approach is advantageous because of the reduction of the number of readout channel, with respect to having to readout one channel per each position. Based on such idea, at FBK (Trento, Italy), a particular type of position-sensitive silicon photomultiplier detector has been developed~\cite{Gola2013ANA}. These devices are called linearly-graded SiPMs (LG-SiPMs). Based on such position sensitive technology, it is possible to have large chips (few squared millimeter, up to 1 squared centimeter) with position resolution. These chips can be also arranged in arrays, having a detection module based on 4 elements of linearly-graded SiPMs (LG-SiPMs). Each element joined based on a charge-sharing architecture integrated on the die (see figure~\ref{fig:schematic_channels}-left) that has a high resolution, low distortion, and fast output signals~\cite{Gola2013ANA, Acerbi2023, Acerbi2024, Gola2020}.  

In this work, we used a tile of 2x2 elements of LG-SiPM chips, each one with area of about 8x8 mm$^2$, previously characterized in \cite{FBK-LG-SiPM}. The tile covers a nominal chip area of almost 16x16 mm$^2$. The output signals from the LG-SiPMs chips are arranged in a proper configuration, through a "smart channel" approach, connecting the central ones together, so that the hit position can be inferred with only 6 readout channels as shown in figure~\ref{fig:schematic_channels}-right~\cite{Acerbi2024}. 

LG-SiPM approach drastically reduces the number of channels to infer the hit position in comparison to an array of SiPMs each having a single readout channel. 
However, electronic defects and non-uniformities can induce non-linearity and worsen the position resolution, reducing the number of distinguishable regions and generating distorted images~\cite{Acerbi2024}.

In this contribution we show a method based on Deep Neural Network (DNN) to minimize these distortion effects thus increasing the granularity of the linearly graded sensor. 
Neural networks have been investigated on Monte Carlo simulation for event positioning in \cite{Jaliparthi2021DeepRN}.

\section{Instrumentation and Methods}

\begin{figure}
    \centering
    \includegraphics[width=.395\linewidth]{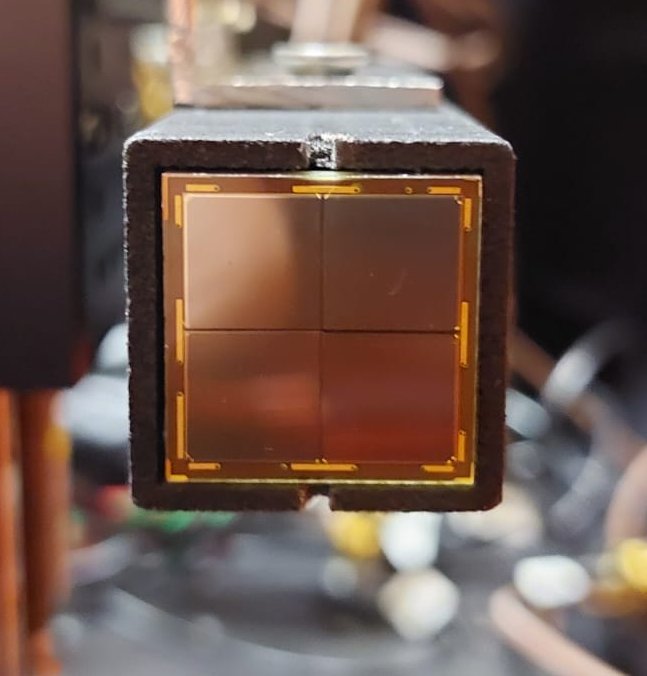}
    \includegraphics[width=.55\linewidth]{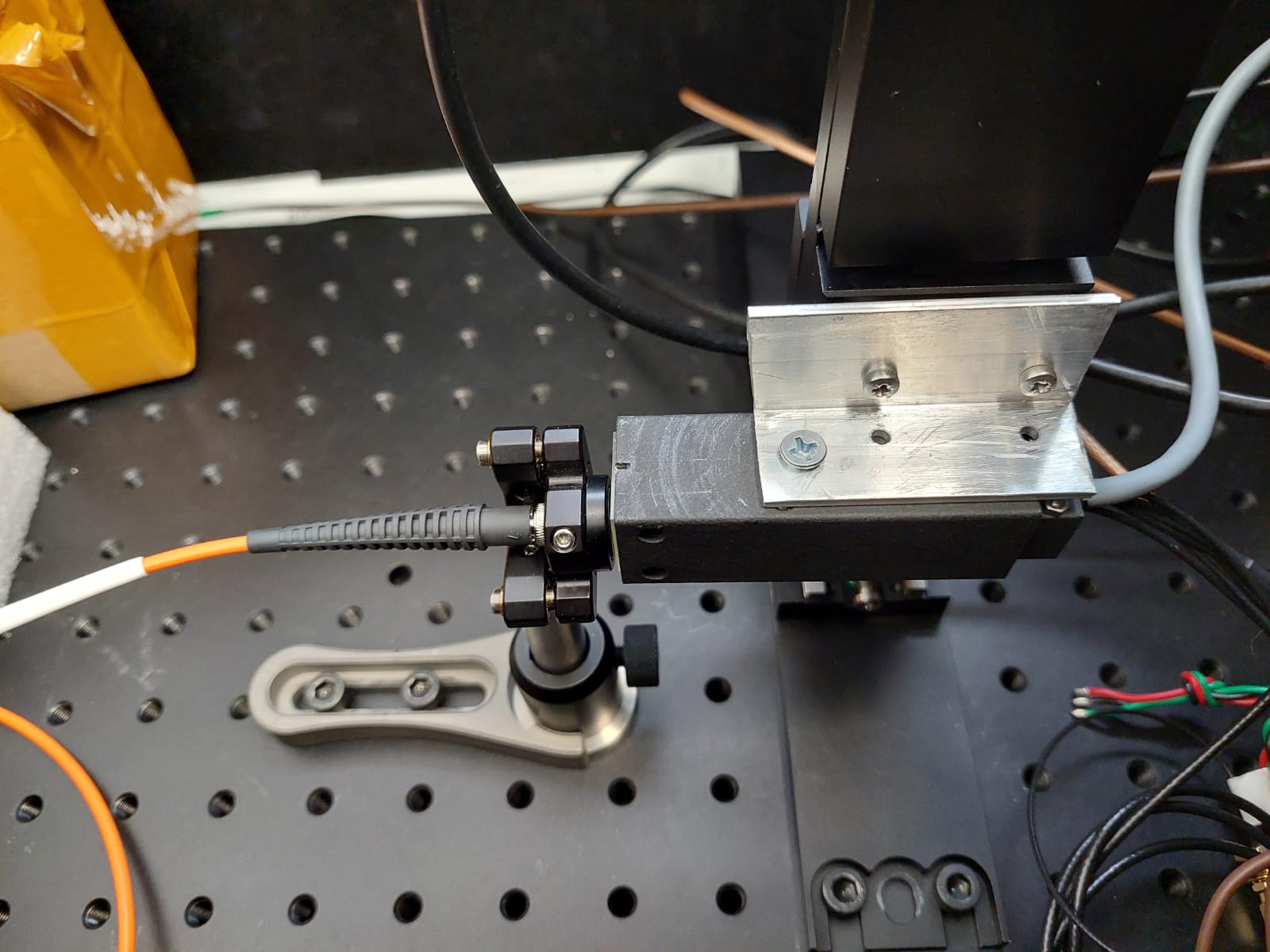}
    \caption{Left: front view of the SiPM tile, mounted in a compact module including signal front-end amplifiers, composed of 4 LG-SiPM. Right: SiPM module mounted on linear stages and with optical fiber placed in front.}
    \label{fig:setup}
\end{figure}

The four element arrays of about 8x8 mm$^2$ LG-SiPMs (see figure~\ref{fig:setup}) are based on the FBK RGB-HD SiPM technology with a $20~\mu\text{m}$ cell pitch. This type of device allows to reach a sub-millimeter position resolution on a nominal 16x16 mm$^2$ chip area.

We biased the LG-SiPM at 33 V (5 V over-voltage) and amplified the signals of the 6 channels. A blue LED (470~nm peak wavelength) is connected to an optical fiber that is placed in front of the LG-SiPM. Although the light spread should not affect the reconstructed position, since LG-SiPMs are designed to reconstruct the center of gravity of the light beam, we placed the fiber as close as possible to limit the spread of the light. We estimated that the light spot size is about 2 mm in diameter. 

We moved the LG-SiPM with two 50mm-range linear stages (see figure~\ref{fig:setup}) 37 steps of 0.5 mm both in the $x$ and $y$ direction. 
At each step, we acquired 10k waveforms per channel of 1000 samples with $800~\rm{ps}$ sampling time with a 1 GHz bandwidth oscilloscope. 

A pulse generator was used to both trigger the oscilloscope and drive the LED. 
The LED was pulsed at 10 kHz and with a square pulse width of 20 ns (corresponding to the shortest pulse that can be achieve with the pulse generator). The voltage amplitude of the square pulse was selected to avoid saturation of the sensor and amplifiers and to maximize the number of photons emitted thus maximizing the precision of the reconstructed position.

The signal amplitude of the 6 LG-SiPM channels ($Q_i,~i=1,~2,~\ldots,~6$) is evaluated and allows to estimate the total charge $Q = \sum_i Q_i$ and reconstruct the hits position ($x_{\mathrm{reco}}, y_{\mathrm{reco}}$) with a standard method and a DNN (see sections \ref{sec:standard} and \ref{sec:dnn}). We then compared the two methods (see section \ref{sec:comp}).

\subsection{Data splitting}\label{sec:splitting}

We split the data set into two samples of the same size, one for training and the other for testing and checking over-fitting. We investigated three splitting techniques : 

 \begin{enumerate}[label=\textbf{(\Alph*)}]
    \item a random split across data set;
    \item a chessboard splitting (selecting the scanned position in a chessboard configuration for training and testing samples);
    \item and a random position splitting (splitting the train and data set between different randomly selected motor coordinates).
\end{enumerate}

\begin{figure}
    \centering
    \includegraphics[width=0.32\linewidth]{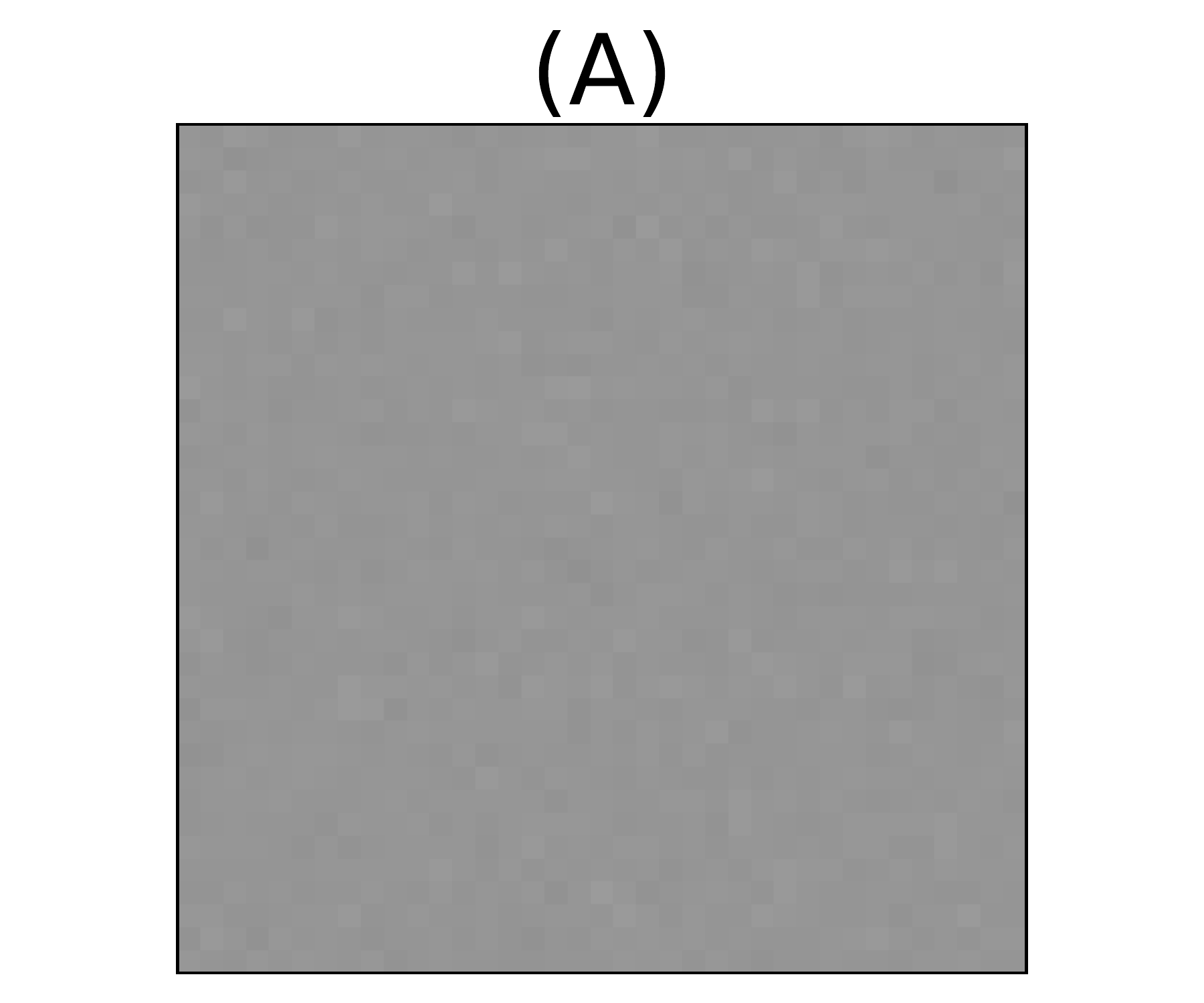} 
    \includegraphics[width=0.32\linewidth]{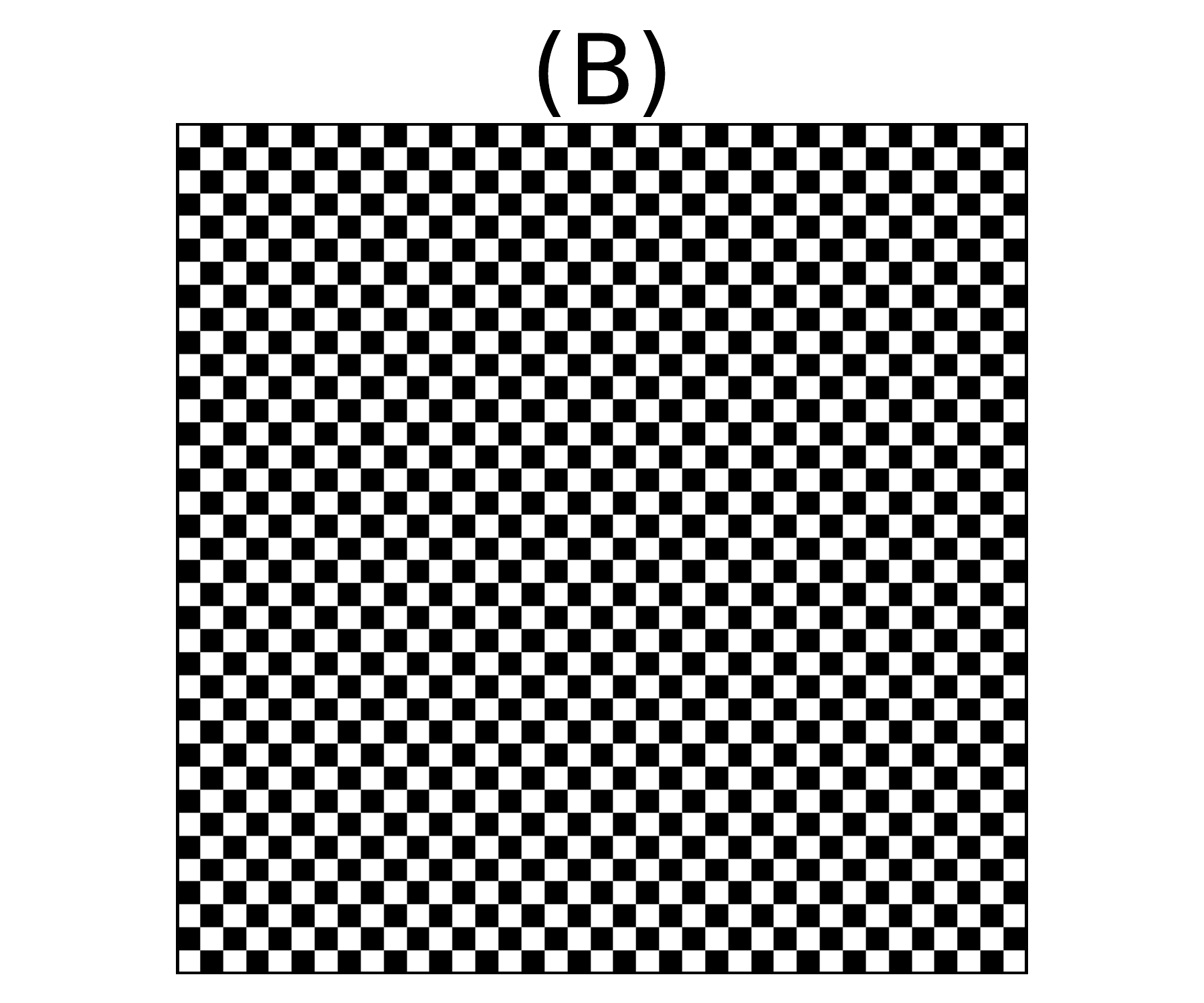}
    \includegraphics[width=0.32\linewidth]{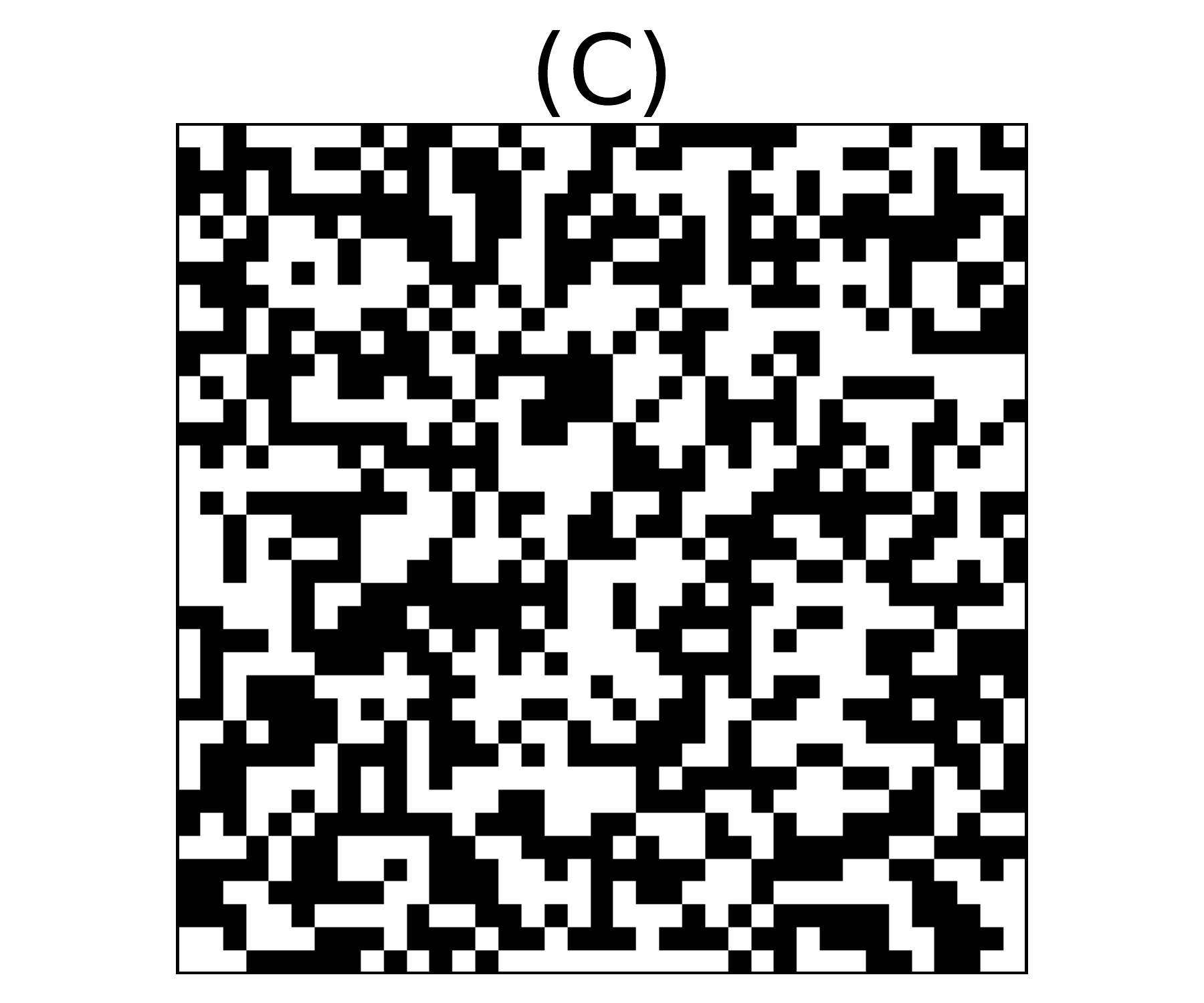}
    \caption{Illustration of the 3 splittings used for training and testing. The gray scale indicates the splitting ratio for each motor position.}
    \label{fig:splitting}
\end{figure}

These three splittings are designed to avoid bias towards the true positions (motor positions) of the training sample which are placed in a regular two-dimensional grid. The three data splitting patterns used are illustrated in figure~\ref{fig:splitting}.

\begin{figure}
    \centering
    \includegraphics[width=0.5\linewidth]{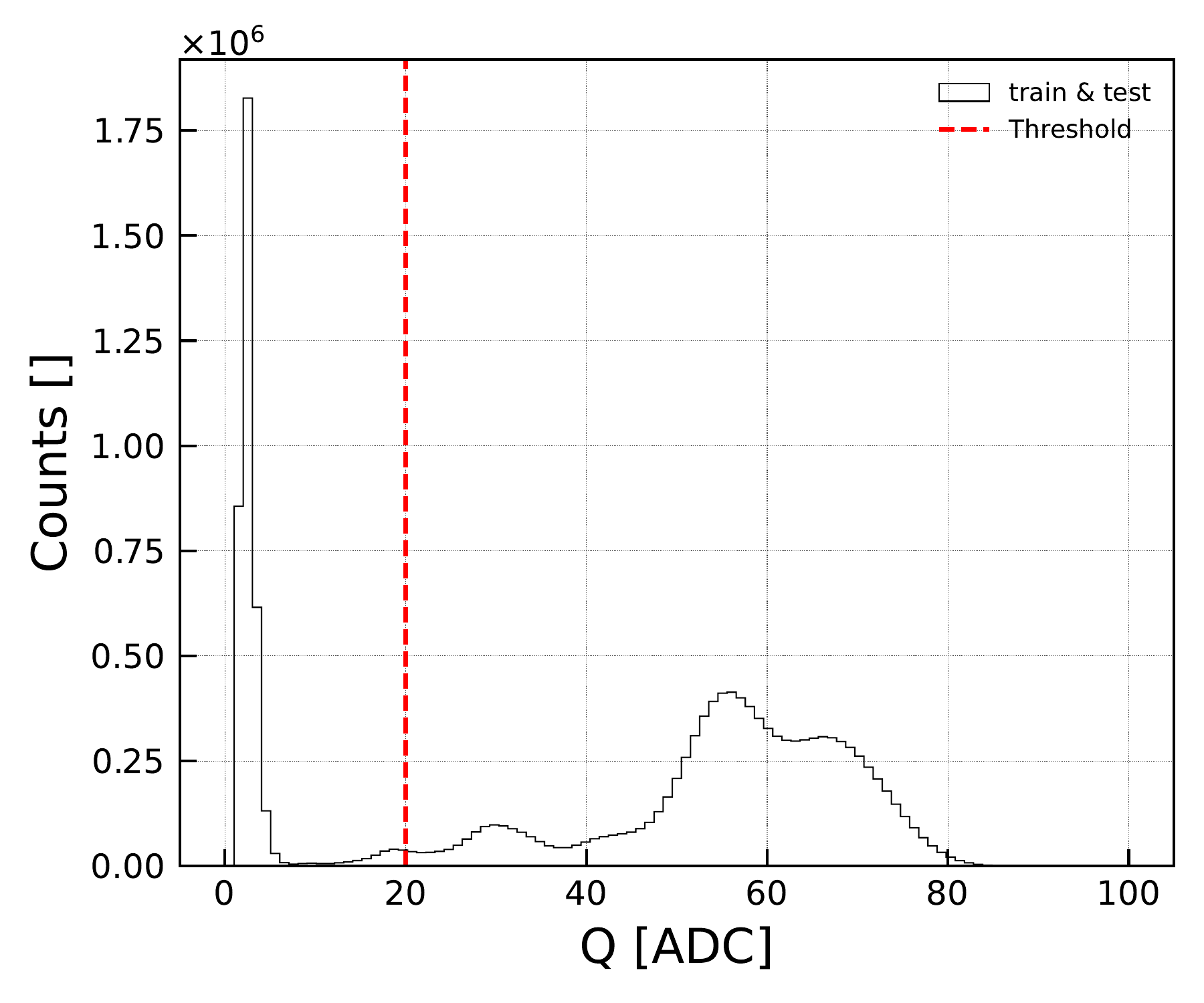}
    \caption{Total charge distribution (black) of the test and train sample. A lower threshold (red) on $Q$ is applied to filter the data.}
    \label{fig:energy-cut}
\end{figure}

After each splitting, we applied a quality cut on the total charge to both samples (see figure~\ref{fig:energy-cut}) to filter out fiber positions that are outside the field of view of the LG-SiPM sensitive area thus removing events falling in the edges of the device, and in the gaps present between SiPM tiles (see figure~\ref{fig:setup}). 

\subsection{Linear reconstruction}\label{sec:standard}

Using the formula for the position reconstruction for an LG-SiPM in \cite{Gola2013ANA} we can derive the relative $x$ and $y$ coordinates for a 2x2 tile of LG-SiPM as:

\begin{eqnarray}\label{eq:relative_coordinates}
    x =& \frac{1}{2} \frac{Q_3 - Q_1}{Q_1 + Q_2 + Q_3}, \\
    y =& \frac{1}{2} \frac{Q_6 - Q_4}{Q_4 + Q_5 + Q_6}.
\end{eqnarray}

The channel number $Q_i$ can be identified from figure~\ref{fig:schematic_channels}-right.

These reconstructed positions are relative to the 2x2 tile of LG-SiPM dimensions. To compare them to the actual coordinates, we need to transform these coordinates in  absolute coordinates (i.e. motor coordinates). To scale these relative coordinates in the linear stage frame we apply scaling factors $l_x$ for $x$ direction and $l_y$ for $y$ direction. The coordinate frames are also shifted and could be slightly tilted by an angle. Thus, the reconstructed positions are then given by the linear applications:

\begin{eqnarray}\label{eq:standard_formula}
    x_{\mathrm{reco}} =& \cos(\phi) \left( l_x  x - x_0 \right) - \sin(\phi) \left(  l_y y - y_0 \right), \\
    y_{\mathrm{reco}} =& \sin(\phi) \left(  l_x  x - x_0 \right) + \cos(\phi) \left( l_y y - y_0 \right),
\end{eqnarray}

where $\phi$ is the tilt angle of the sensor, $x_0$ and $y_0$ are the central coordinates of the LG-SiPM with respect to the linear stages coordinate system. $l_x$ and $l_y$ are the effective height and width of the LG-SiPM. 

These parameters can be computed on the training sample by minimizing the average distance squared between the reconstructed positions and the motor positions.

\subsection{Neural Network reconstruction}

A more advanced approach to the simple linear reconstruction in section~\ref{sec:standard}, that follows from the detector geometry, is based on a neural network architecture. 

Here we first introduce a "zero hidden" layer to motivate the use of a neural network later in section~\ref{sec:zero-dnn}. We then increase the number of hidden layers in \ref{sec:dnn}.

The performance of the DNN for an increased number of layers is shown in section~\ref{sec:comp}.

\subsubsection{Zero Hidden Layers Neural Network}\label{sec:zero-dnn}

The standard formula (see equation~\ref{eq:standard_formula}) for reconstructing the position $x_{reco}$ and $y_{reco}$ can be rewritten as follows:

\begin{equation}\label{eq:linear_system}
    \begin{pmatrix}
    x_{{\rm reco}} \\
    y_{{\rm reco}}
    \end{pmatrix} = R(\phi) \left(A \vec{Q} - \vec{b} \right),
\end{equation}

with $R(\phi)= \begin{pmatrix}
    \cos \phi & -\sin \phi \\
    \sin \phi & \cos \phi
\end{pmatrix}$ the rotation matrix, $\vec{Q} = \left( Q_1, Q_2, \ldots, Q_6 \right)$ the charge vector, $A$ a $2\times 6$ matrix:

\begin{equation}
    A = \frac{1}{2}\begin{pmatrix}
        -l_x & 0 & l_x & 0 & 0 & 0 \\
        0 & 0 & 0 & -l_y & 0 & l_y \\
    \end{pmatrix},
\end{equation}

and $\vec{b} = \left(x_0, y_0 \right)$ the shift vector. 

By design of the sensor, some elements $A_{ij}$ of the matrix $A$ should be equal to zero. Here, we propose a model where we introduce all 12 parameters of the $A$ matrix.

In practice, the linear system of equations~\ref{eq:linear_system} can be rewritten in the form $A^{\prime} \vec{Q} - \vec{b^\prime}$ representing a linear model with 14 free parameters. Which is a "zero hidden" layer neural network with 6-unit input layer with linear activation function towards towards a 2-output layer. 

\subsubsection{Deep Neural Network}\label{sec:dnn}

The "standard" formula \ref{sec:standard} assumes a perfectly linear response of each single SiPM cell and an equal gain across the 4 tiles. This model is simply a system of to 2 linear equations with 5 free parameters. 

The "zero hidden" layer model in section~\ref{sec:zero-dnn} adds more free parameters in order to account for the importance of each channel to the other. It can be interpreted as a weighted computation of the relative coordinates from equations~\ref{eq:relative_coordinates}. In this case it corresponds to a linear application with 14 free parameters. Assuming that the LG-SiPM response is linear the weighting parameters of the 6 channels should be -1, 1 or 0. Thus adding the weighting parameters should allow to better parameterize the sensor response.

To compensate for non-linearities, the model itself should be non-linear. We do this by increasing the number of layers and using non-linear hyperbolic tangent activation functions. The number of units per layer is also increased thus adding more free parameters that captures the sensor non-linear response.

\begin{figure}
    \centering
    \includegraphics[width=\linewidth]{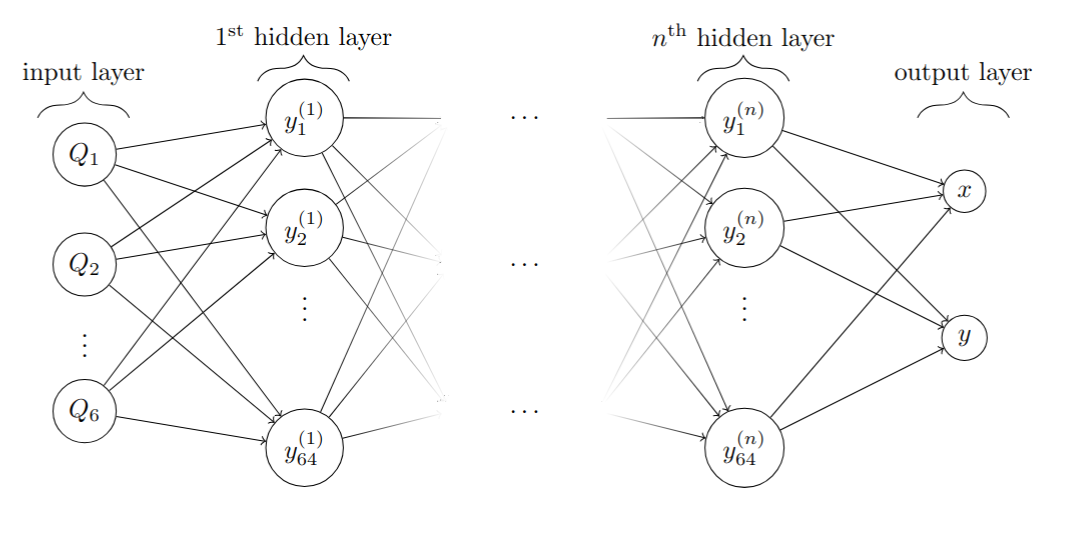}
    \caption{Schematic of the Deep Neural Network architecture used. The DNNs consist of an input layer for the amplitude $Q_i$ connected to a series of $N_{\rm{layers}}$ 64-input dense hidden layers with hyperbolic tangent activation functions. The output layer is connected to the last hidden layer by a linear activation function to reconstruct the coordinates.}
    \label{fig:dnn}
\end{figure}

We built several deep neural networks all having a 6-unit input layer (one for each channel amplitude $Q_i$) connected to a series of $N_{\rm{layers}}$ hidden dense neural network layers with 64 units per layer and hyperbolic tangent activation function (see figure~\ref{fig:dnn}). 

The dataset was divided into independent training and testing samples as explained in~\ref{sec:splitting}. The training sample was used to optimize the network weights and biases, while the validation sample was used to evaluate the performance of the networks (see section~\ref{sec:comp}).

Before training, the input $Q_i$ were normalized by the total collected charge in order to remove the dependence on the absolute light yield:
\begin{equation}\label{eq:normalization}
    Q_i \rightarrow \frac{Q_i}{\sum_{i=1}^{6} Q_i}. 
\end{equation}

This normalization stabilizes the training and ensures that the network learns from the relative charge distribution rather than from the absolute signal amplitudes.

The output layer consists of 2 units dense layer with a linear activation for the $x_{\mathrm{reco}}$ and $y_{\mathrm{reco}}$ reconstructed positions.

The loss function was set as the mean squared error between the reconstructed positions ($x_{\mathrm{reco}}, y_{\mathrm{reco}}$) and motor positions ($x_{\mathrm{motor}}, y_{\mathrm{motor}}$).

The networks were trained in batches during 40 epochs using the Adam optimizer~\cite{Adam} with an initial learning rate $\alpha = 10^{-3}$, exponential decay rates $\beta_1 = 0.9$, $\beta_2 = 0.999$ and $\epsilon = 10^{-7}$. 

The model weights were initialized using the Glorot uniform initialization, which is well suited for hyperbolic tangent activation functions and helps maintain stable gradients during training~\cite{Glorot2010}. The model biases were initialized to zero.

Training convergence was verified by monitoring both the training and validation losses as a function of epoch number (see section~\ref{sec:loss_epoch}).

All models were implemented using the \textit{Keras} deep learning API written in Python \cite{chollet2015keras}. 

\section{Results}
\label{sec:comp}
\subsection{Qualitative results}

\begin{figure}
    \centering
    \includegraphics[width=0.31\linewidth]{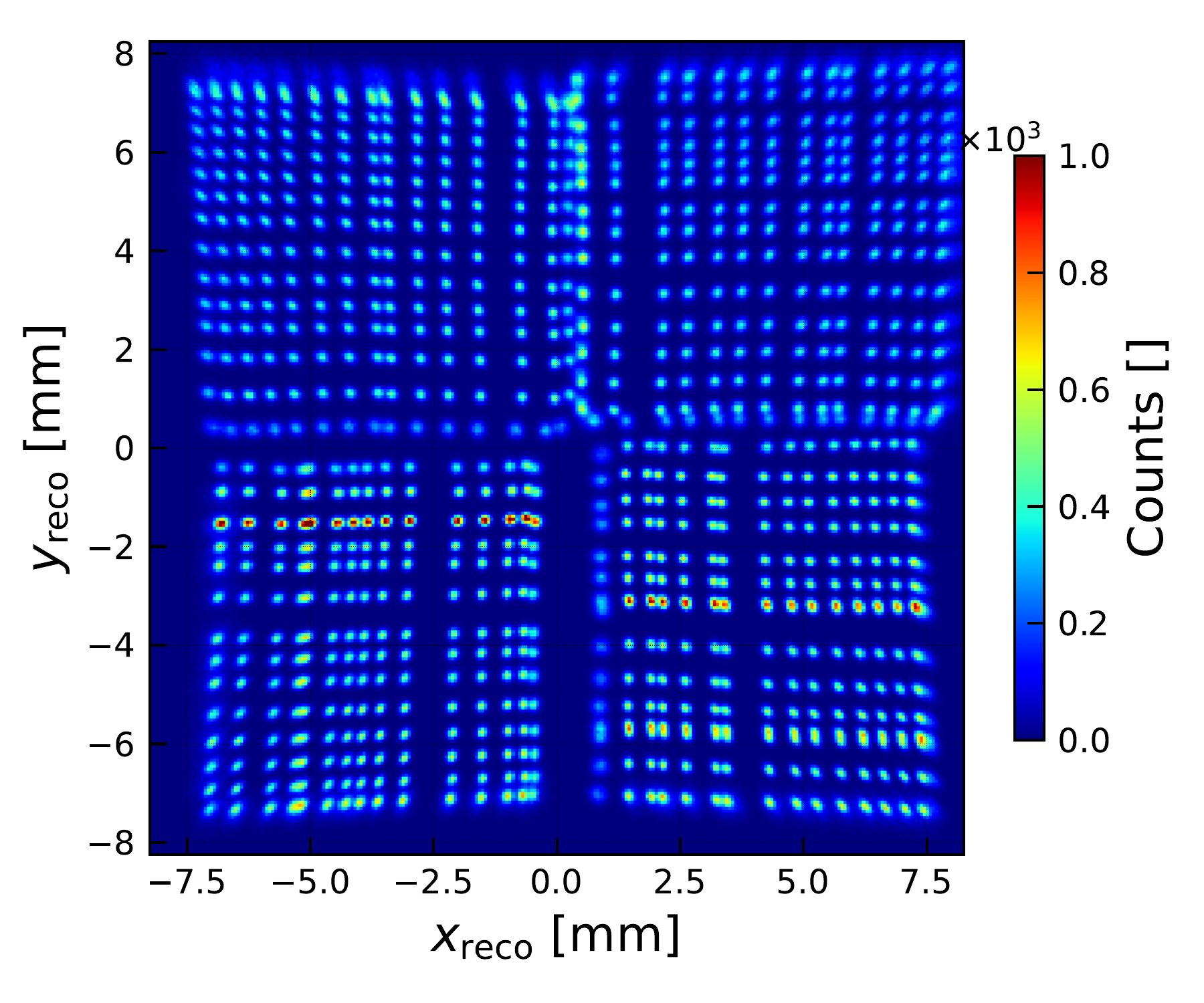}
    \includegraphics[width=0.31\linewidth]{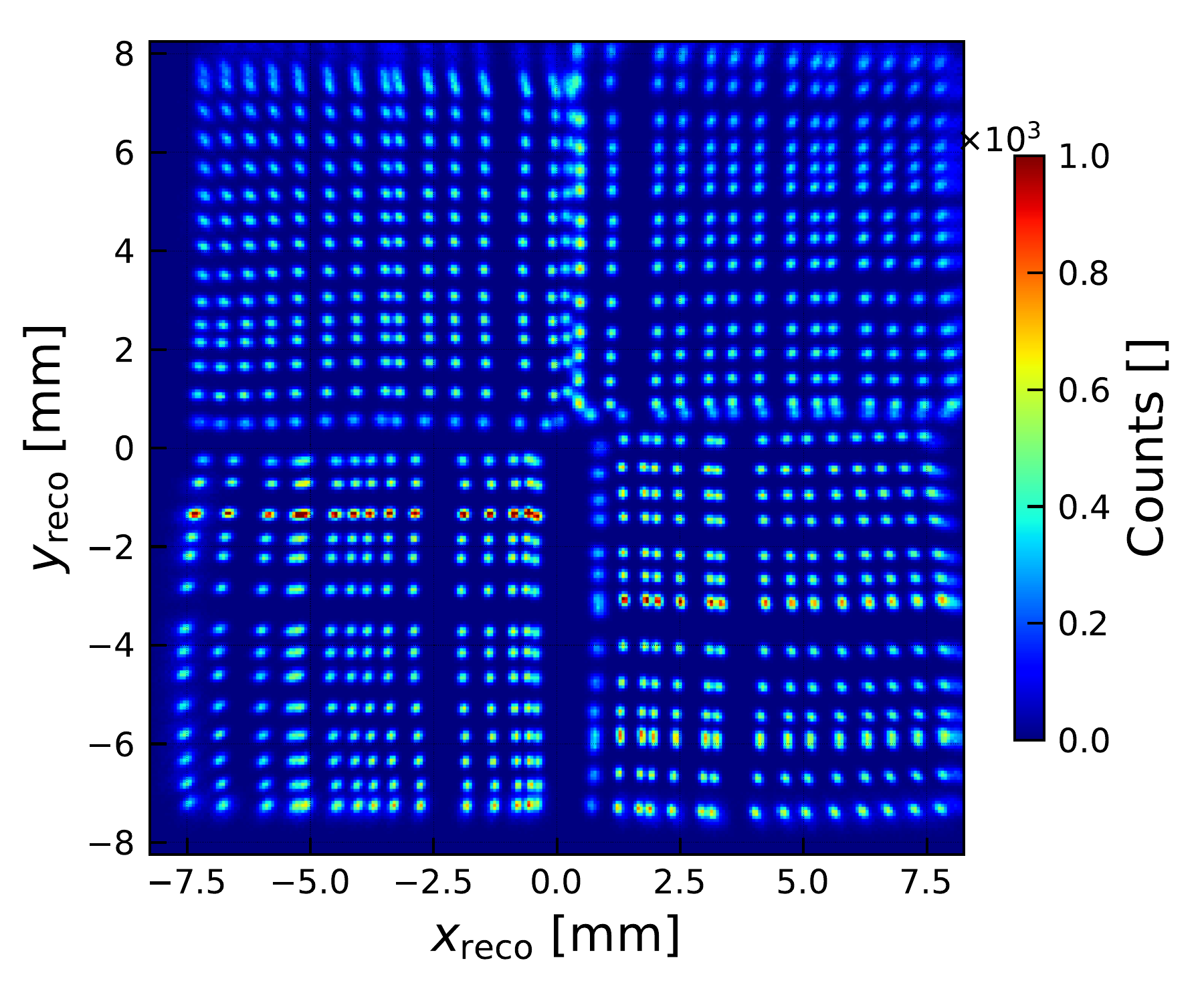}
    \includegraphics[width=0.31\linewidth]{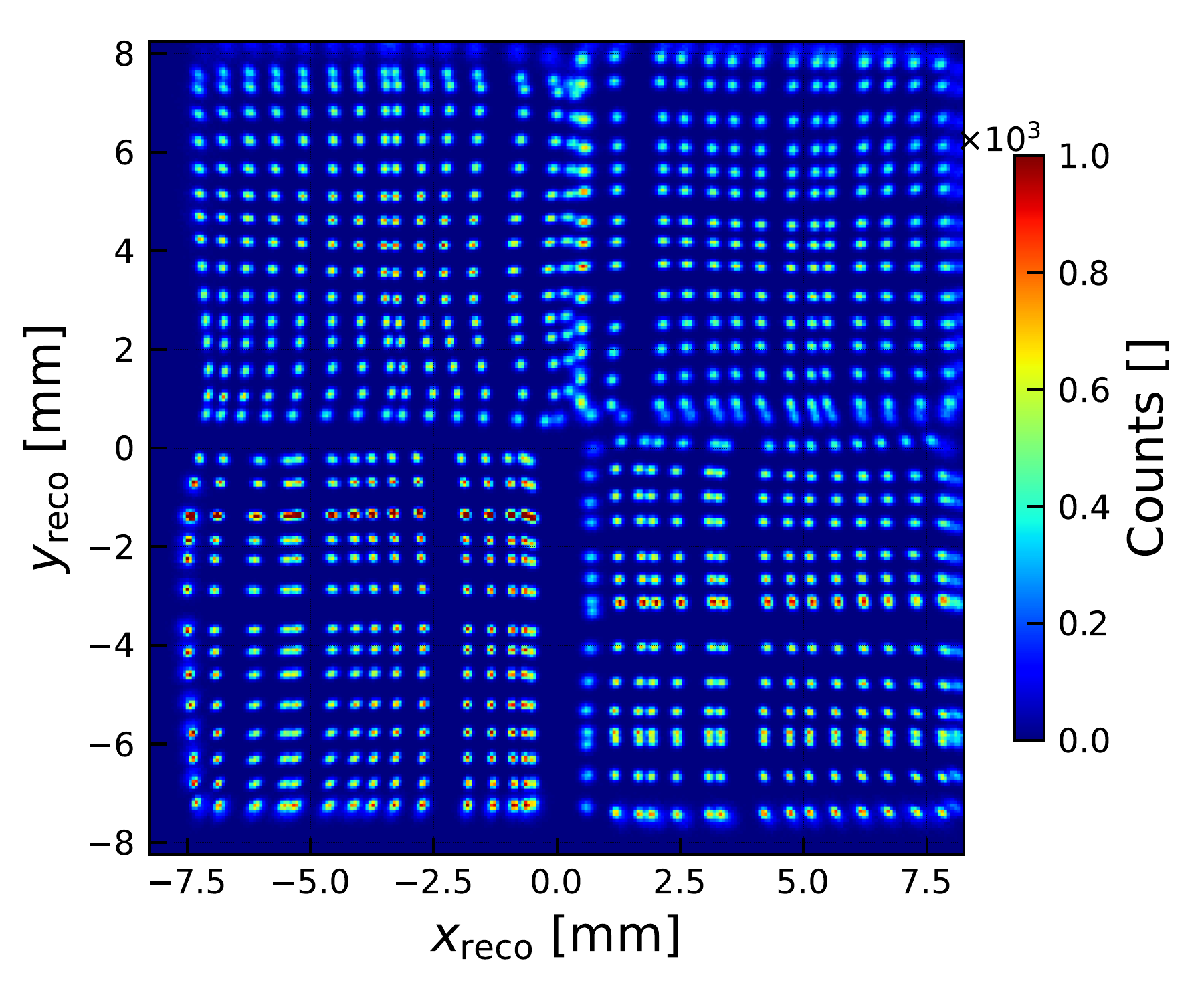}
    \includegraphics[width=0.31\linewidth]{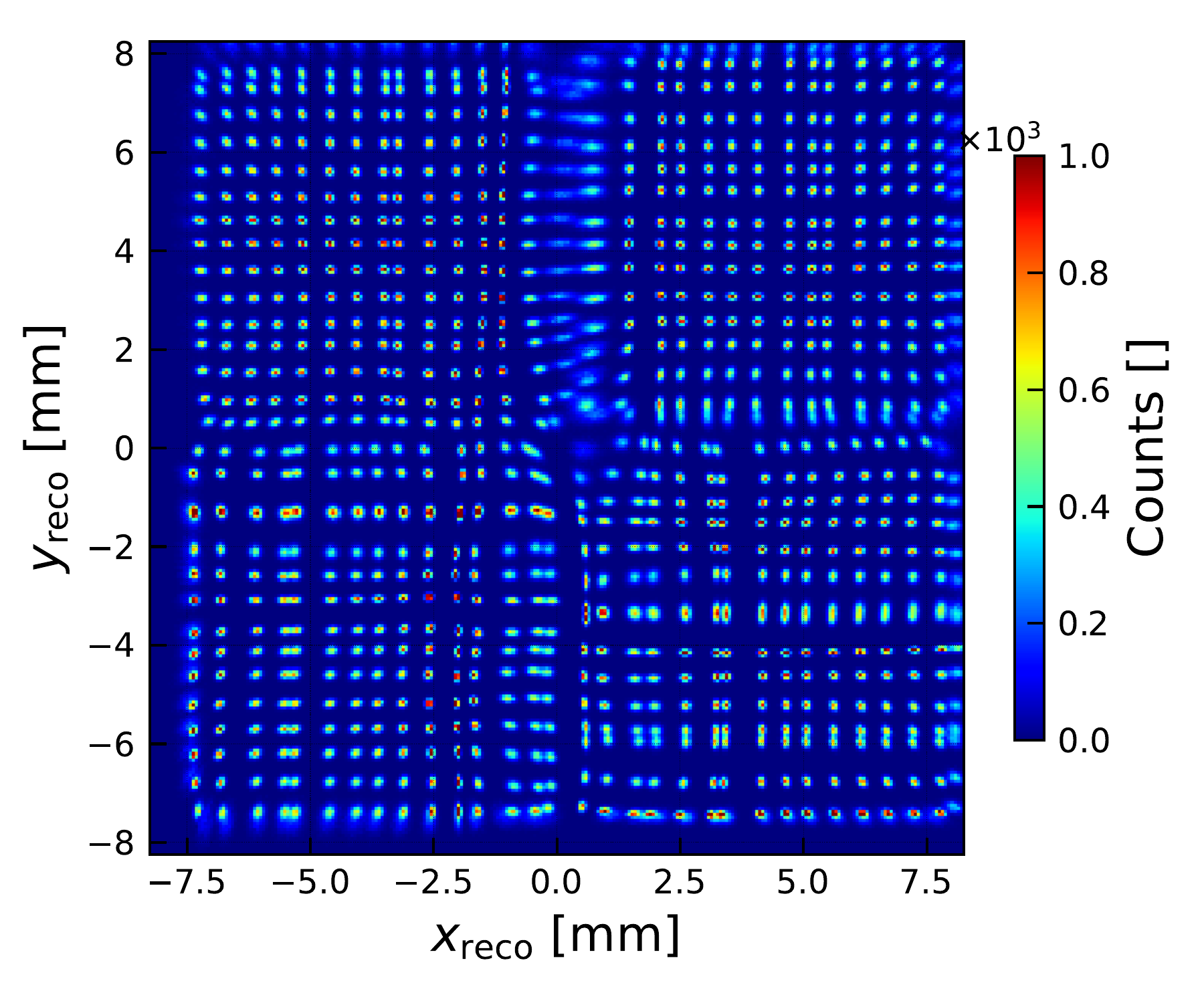}
    \includegraphics[width=0.31\linewidth]{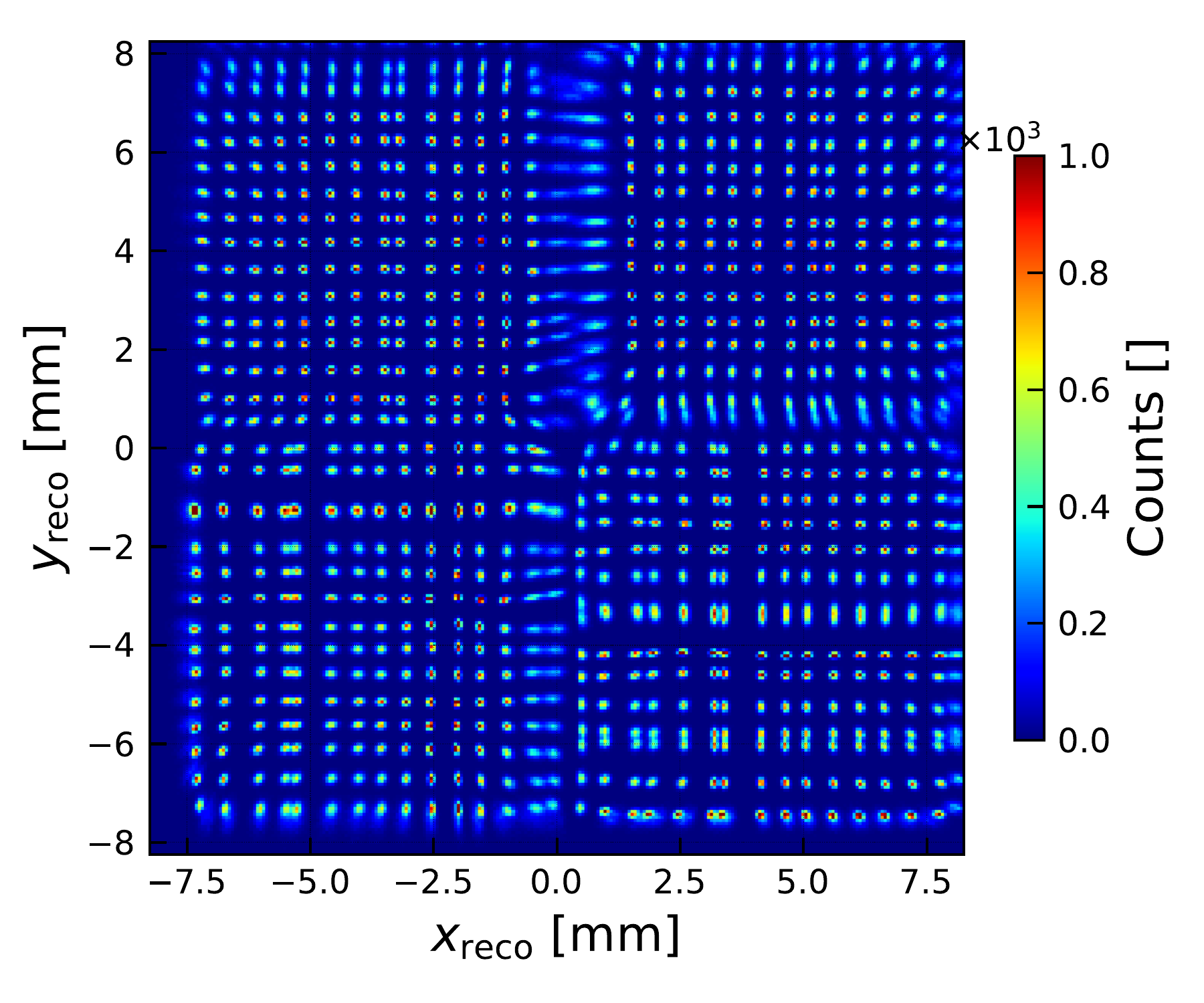}
    \includegraphics[width=0.31\linewidth]{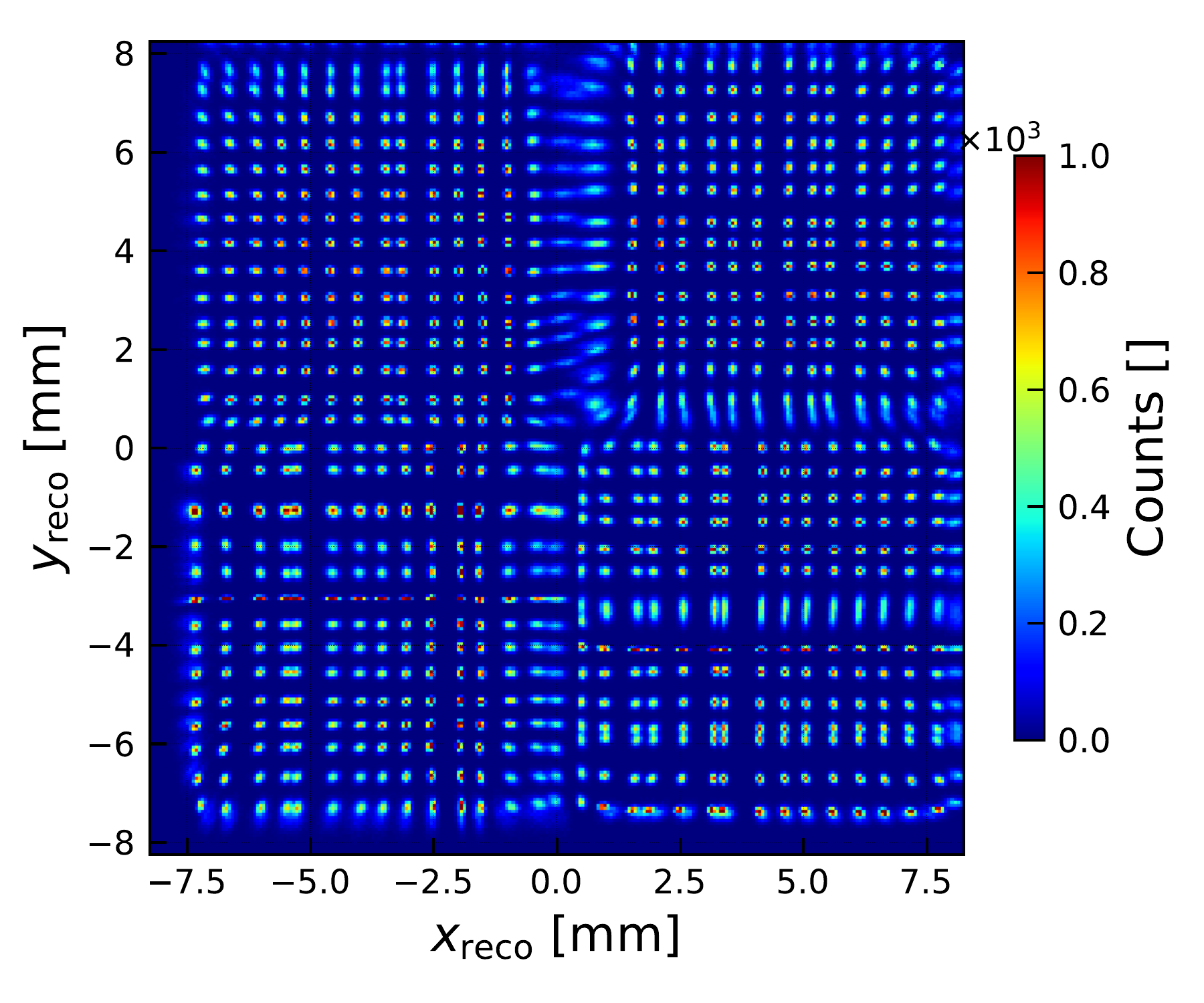}
    \caption{Reconstructed position of the test sample for the splitting technique \textbf{(A)}. Each image corresponds to a number of layers used in the DNN (in reading order: $N_{\rm{layers}} = 0,~1,~\ldots,~5$).}
    \label{fig:image_reco_layers}
\end{figure}

Reconstructed images of the scan are shown in figure~\ref{fig:image_reco_layers} for the splitting technique \textbf{(A)} on the test sample. Each image represents a different number of layers $N_{{\rm layers}}$ used. One can see that the non-linearities are recovered by the DNN. The reconstructed images from the linear reconstruction presented in section~\ref{sec:standard} can be found in~\cite{raiola2025spatialresolution}. 

Large position reconstruction shift can be observed at the gaps of the SiPM tiles, borders of the LG-SiPMs and in some rows or columns of particular SiPM tiles systematically. Although these discrepancies are not completely overcome by the DNN we observe that the reconstruction error is improved. These systematic shift in the position errors could be explained by the imperfections in the linearly-graded resistive network or the quality cut which inherently cuts events at the borders and in the gaps thus reducing sample to train events in these regions.   

Quantitative assessment of the DNNs is further discussed in sections~\ref{sec:result_reso} and \ref{sec:result_granu}. 

\subsection{Loss as function of epoch}\label{sec:loss_epoch}

\begin{figure}
    \centering
    \includegraphics[width=0.32\linewidth]{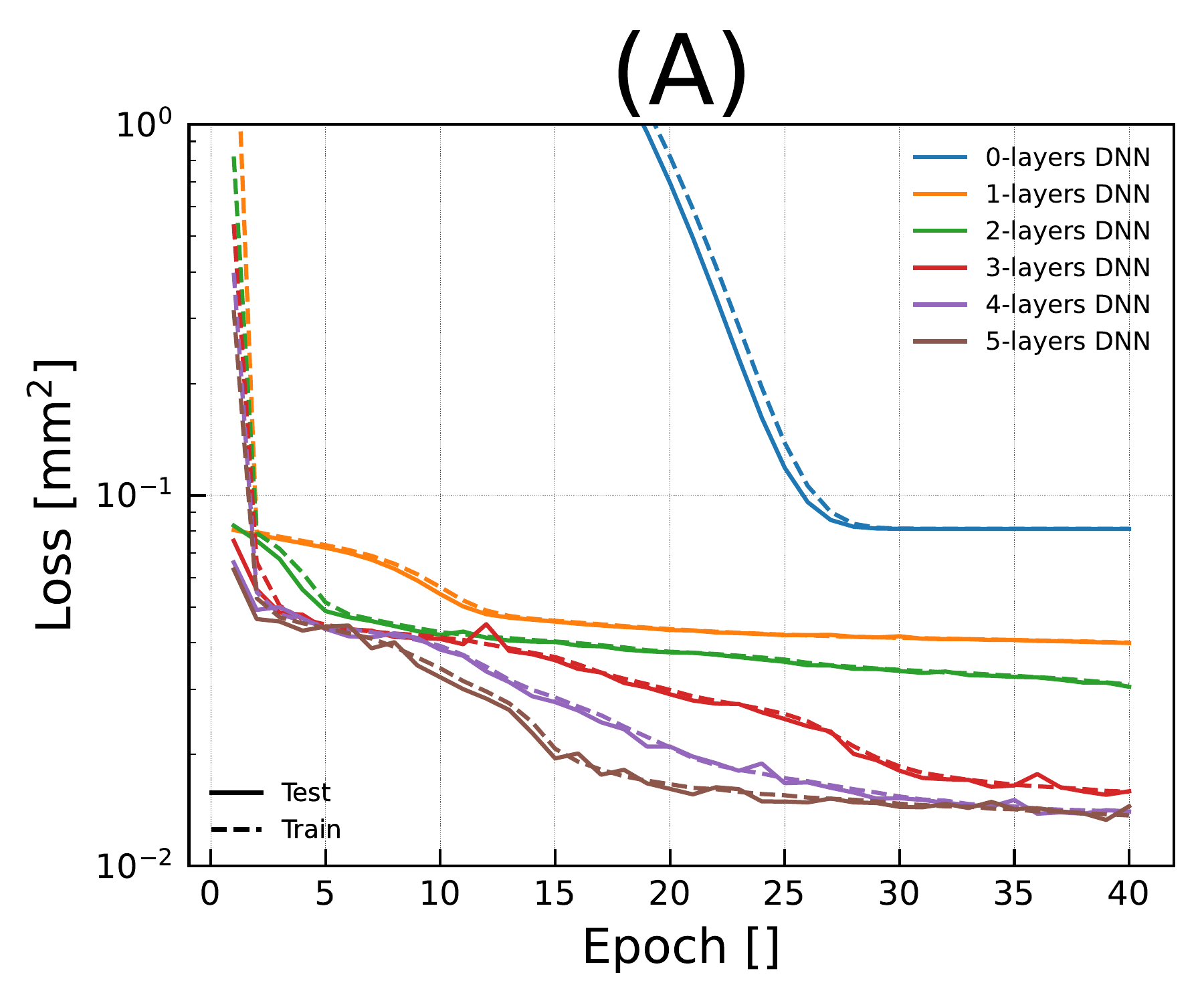}
    \includegraphics[width=0.32\linewidth]{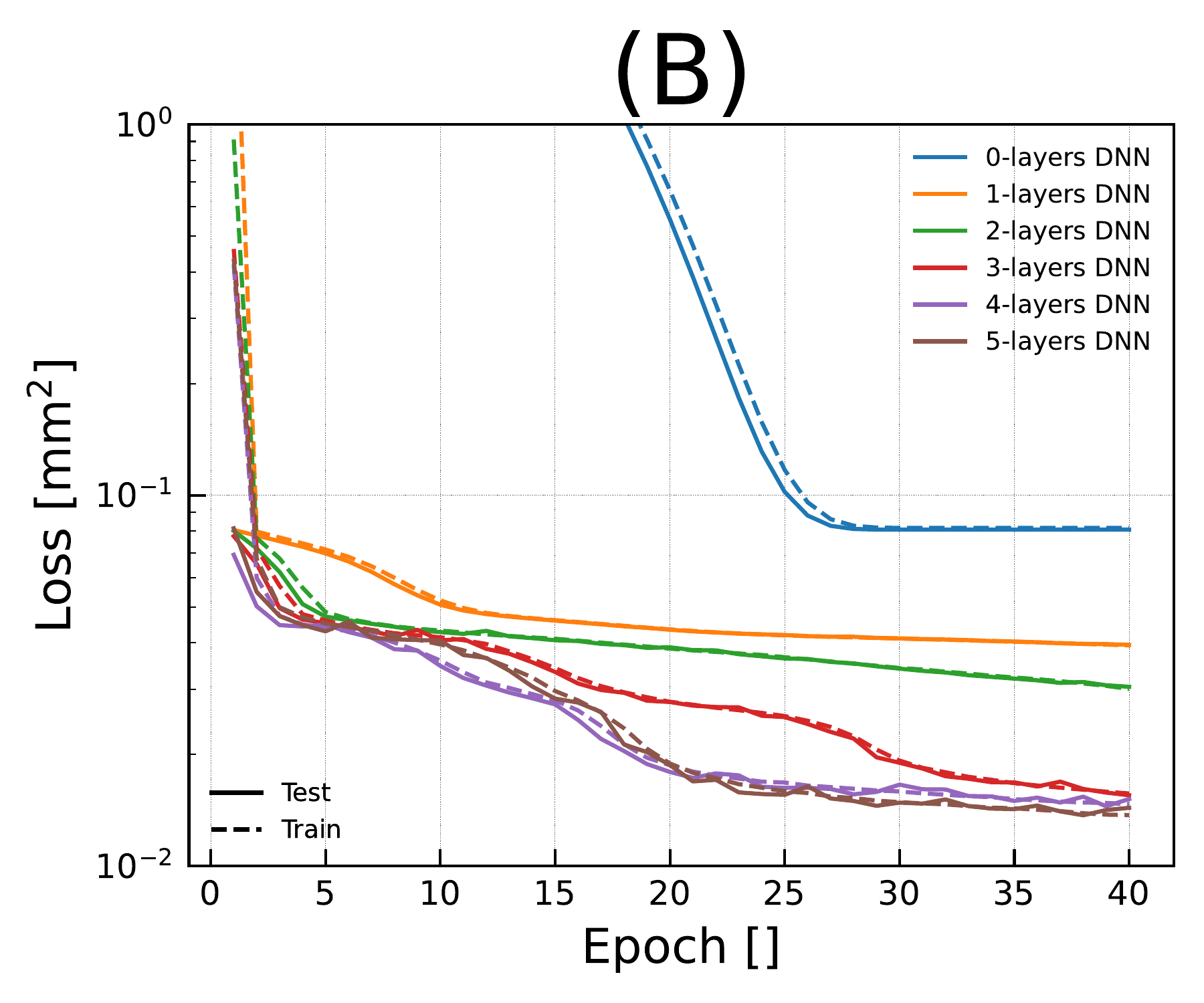}
    \includegraphics[width=0.32\linewidth]{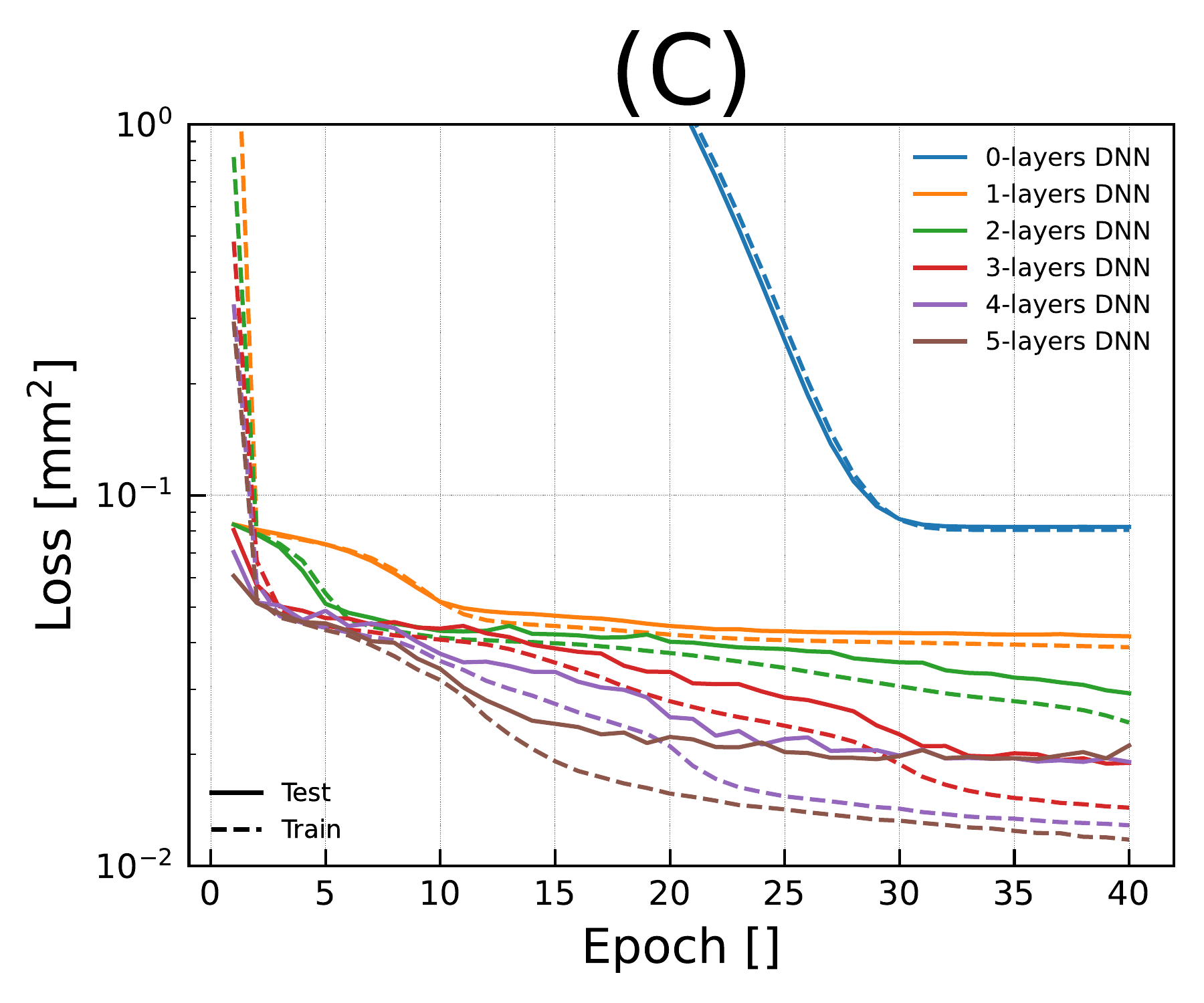}
    \caption{Loss of each DNN during training as function of the epoch for the training sample (dashed lines) and the testing sample (solid lines). The loss is shown for each splitting technique \textbf{(A)} (left), \textbf{(B)} (center), \textbf{(C)} (right).}
    \label{fig:history_dnn}
\end{figure}

The loss as function of the training epoch are shown in figure~\ref{fig:history_dnn} for each splitting techniques and for all neural networks. One can see that the loss of the zero hidden layer model (linear model) is slowly converging towards a minimum (after 25 epochs) while the non-linear models (with more than 1 hidden layer) rapidly surpass the linear model (after about 3 epochs). Increasing the number of hidden layers above 3 does not drastically improve the loss at the end of the training. Whereas having 1 or 2 hidden layers is not optimum.

\subsection{Resolution and systematic shift}\label{sec:result_reso}

To compare the neural networks performances, we computed the systematic shift and resolution of the methods at each scanned position of the test sample. For each step, we computed the Euclidean distance $d$ between the reconstructed and motor position:

\begin{equation}
    d = \sqrt{(x_{\mathrm{reco}} - x_{\mathrm{motor}})^2 + (y_{\mathrm{reco}} - y_{\mathrm{motor}})^2}
\end{equation}

This allows us to evaluate the resolution $\sigma$ (how spread are the reconstructed positions) and average shift $\nu$ of the tested positions (how far is the centroid of the reconstructed positions from the true position). These values were obtained by fitting the distances $d$ to a Rice distribution~\cite{rice} as in~\cite{raiola2025spatialresolution}. The resolution $\sigma$ represents the noise in the detector that cannot be overcome while the shift $\nu$ is a measure of the displacement that occurs between the true and reconstructed positions that can be learned by the DNNs. 

\begin{figure}
    \centering
    \includegraphics[width=0.32\linewidth]{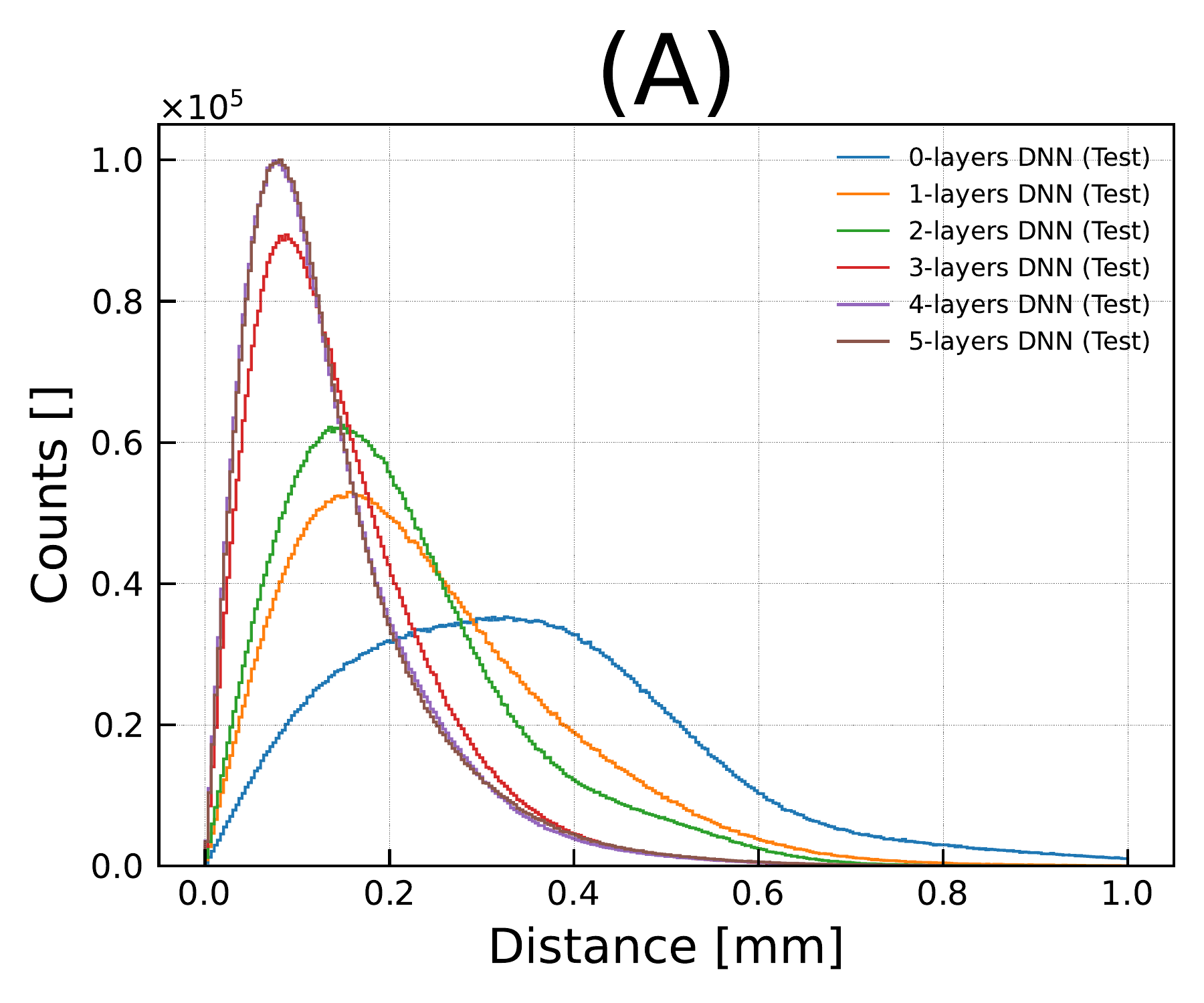}
    \includegraphics[width=0.32\linewidth]{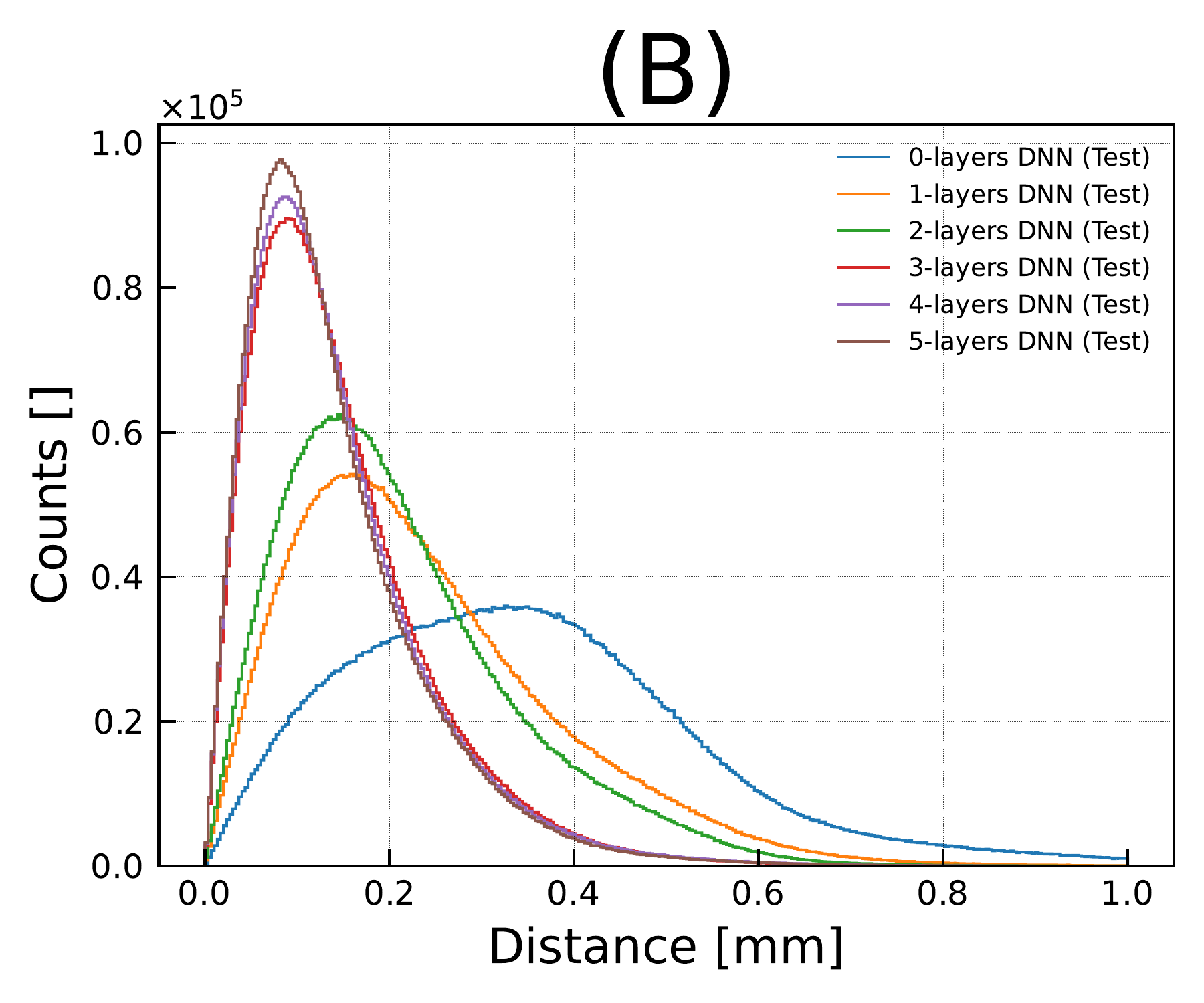}
    \includegraphics[width=0.32\linewidth]{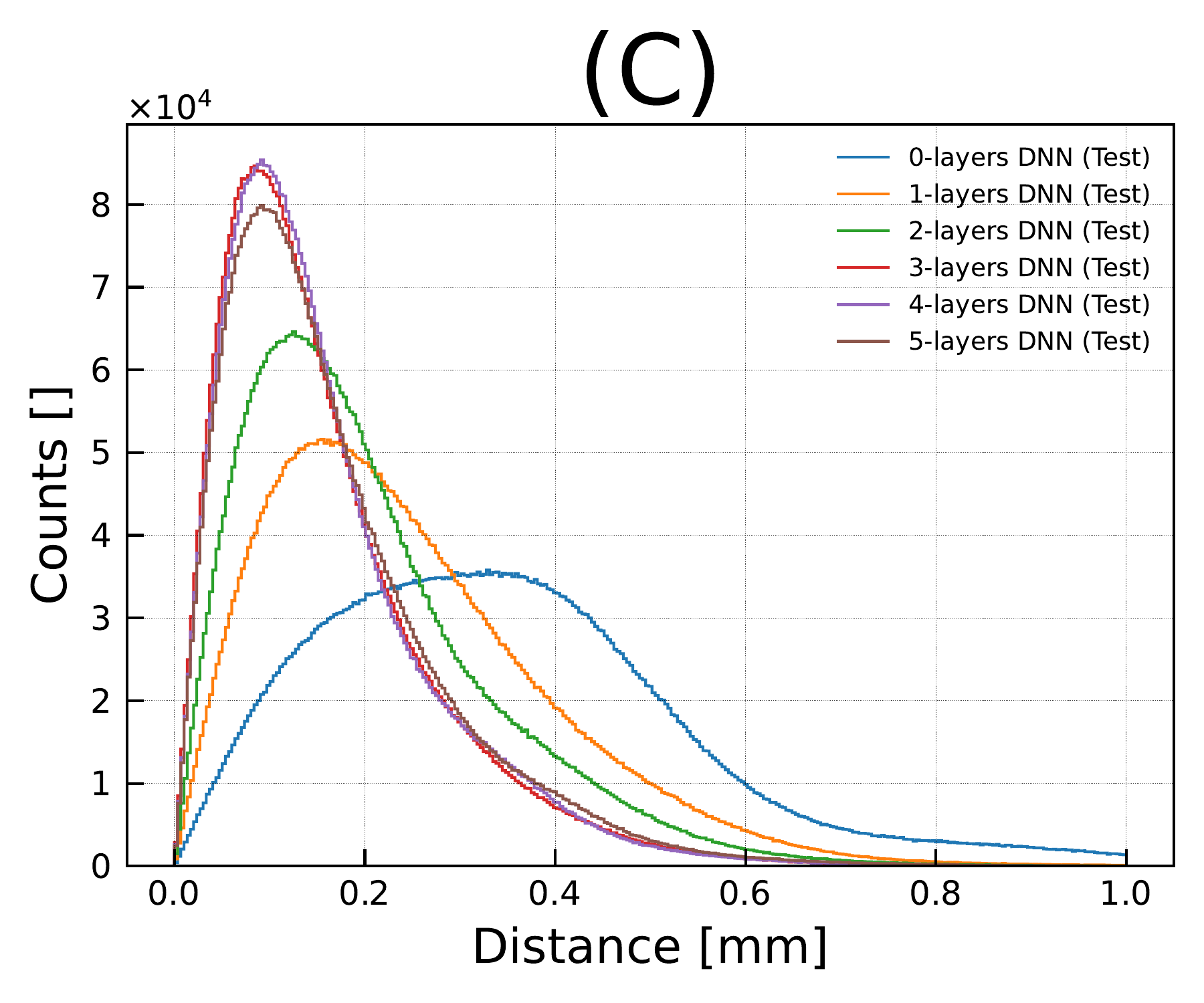}
    \caption{Distribution of the distance $d$ between the reconstructed positions and motor positions for the random splitting \textbf{(A)} (left), the chessboard splitting \textbf{(B)} (center) and the run splitting \textbf{(C)} (right) techniques on the test sample. The distributions are shown for all 0 to 5 layers DNN trained.}
    \label{fig:distance_comparison}
\end{figure}

The results of shift $\nu$, the resolution $\sigma$, and the granularity are presented in table~\ref{tab:results}. We see that in terms of average resolution, the two methods are comparable with $\sigma$ between 66~$\mu$m and 79~$\mu$m. This can be explained by the noise of the device that cannot be completely removed. We attribute this slight improvement to the fact that the non-linear models can locally compensate for distortions in the relative coordinate space, whereas the linear model can only perform a global reshaping of the relative coordinate space (through the $l_x$ and $l_y$ parameters). 
However, the DNN outperforms the linear reconstruction in terms of systematic shift $\nu$ where the DNN is 3.4 to 7.8 times better than the nominal reconstruction technique. This can be also seen in figure~\ref{fig:image_reco_layers} and in figure~\ref{fig:distance_comparison} where we observe that the non-linearities are well corrected by the DNN. 

\subsection{Granularity}\label{sec:result_granu}

Another factor of interest is the granularity which relates to the minimum size of distinguishable areas. This quantity estimates the average size of a distinguishable region in the sensor. It is computed as twice the mean of the distance distribution $2\left< d \right>$ which corresponds to when two neighboring regions are overlapping and thus are indistinguishable. Results of the granularity are reported in table~\ref{tab:results}.

\subsection{Number of resolvable regions}

Additionally to the granularity, one can compute the effective number of distinguishable regions for each method investigated. This number is given by the ratio of the device sensitive area $A = 16 \times 16~{\rm mm^2}$ to the granularity squared:

\begin{equation}
    N = \frac{A}{4 \left< d \right>^2}.
\end{equation}

We observe that the number of distinguishable region is up to $12.1$ times greater from the DNN model compared to the linear model when considering the splitting \textbf{(A)}. For the other splitting methods the improvement factors are $9$ (splitting \textbf{(B)}) and $5.7$ (splitting \textbf{(C)})  In other words, for a field of view of 16x16 mm$^2$, the performances observed for the DNN correspond to a maximum of $6530$ distinguishable regions while for the standard reconstruction, this corresponds to a maximum of $541.5$ distinguishable regions.

\begin{table}
    \centering
    \begin{tabular}{|p{0.35\linewidth}|p{0.15\linewidth}|p{0.15\linewidth}|p{0.15\linewidth}|}
    \hline
        \multicolumn{4}{|c|}{\textbf{Linear model}} \\  \hline
         \textbf{Splitting} & \textbf{A} & \textbf{B} & \textbf{C}\\  \hline
         Mean resolution $\left< \sigma \right>$ \hfill [$\mu$m] & $78.5\pm 0.5$& $78.4\pm 0.5$& $79.3\pm 0.5$\\ \hline
         Mean shift $\left< \nu \right>$ \hfill [$\mu$m] &$317.4\pm 5.6$ & $317.5\pm 5.6$& $311.4\pm5.4$\\\hline
         Granularity $2\left< d \right>$ \hfill [$\mu$m] & $687.6\pm 0.2$& $686.5\pm 0.2$ & $674.8\pm 0.2$  \\ \hline
          N. of resolvable regions $N$ \hfill & $541.5\pm 0.3$ & $543.2 \pm 0.3$ & $562.2\pm 0.3$ \\ \hline
    \end{tabular}

    \begin{tabular}{|p{0.35\linewidth}|p{0.15\linewidth}|p{0.15\linewidth}|p{0.15\linewidth}|}
             \multicolumn{4}{c}{} \\ 
    \hline
         \multicolumn{4}{|c|}{\textbf{DNN model}} \\  \hline
         \textbf{Splitting}  & \textbf{A} & \textbf{B} & \textbf{C}\\  \hline
         Mean resolution $\left< \sigma \right>$ \hfill [$\mu$m] & $67.4\pm1.0$ & $67.7\pm1.3$ & $66.1\pm1.6$ \\ \hline
         Mean shift $\left< \nu \right>$ \hfill [$\mu$m] & $40.9\pm2.6$ & $59.0\pm3.2$ & $92.8\pm3.9$ \\\hline
         Granularity $2\left< d \right>$ \hfill [$\mu$m] & $198.0\pm 0.1$ & $228.7\pm 0.1$ & $283.6\pm 0.1$ \\ \hline
         N. of resolvable regions $N$ \hfill & $6530.0\pm 6.6$ & $4894.0\pm 4.3$ & $3182.9\pm 2.2$ \\ \hline

    \end{tabular} 
        \caption{Mean resolution $\left< \sigma \right>$, mean shift $\left< \nu \right>$, granularity and number of resolvable regions of the linear model (top) and DNN (bottom) for the different splitting between the test and train samples. The averaged values were computed among the scanned positions removing the outer rows and columns. The errors correspond to the statistical errors. The three splittings are: random splitting (A), chessboard splitting (B), and random position splitting (C).}\label{tab:results}
\end{table}


\begin{figure}
    \centering
    \includegraphics[width=0.32\linewidth]{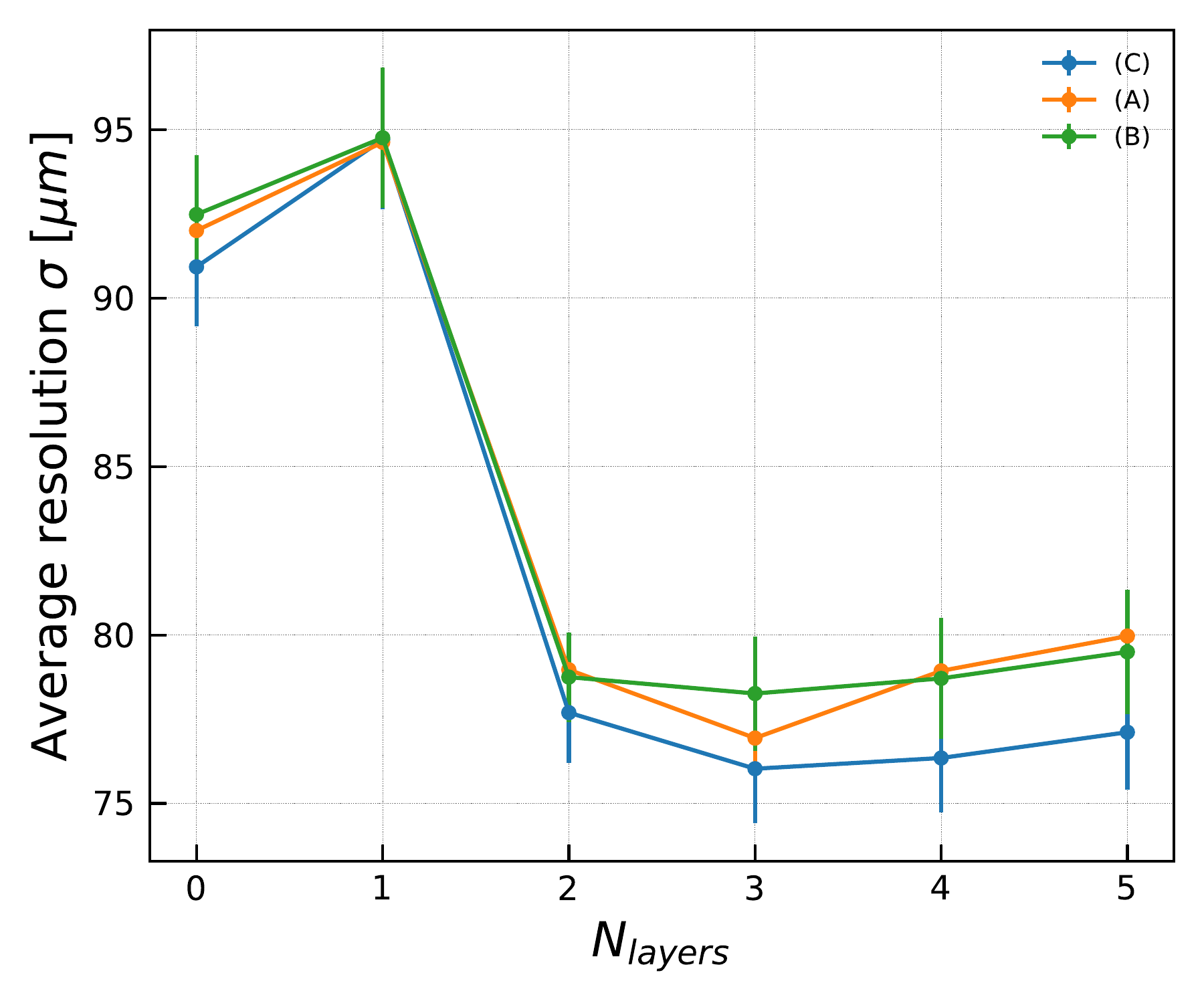}
    \includegraphics[width=0.32\linewidth]{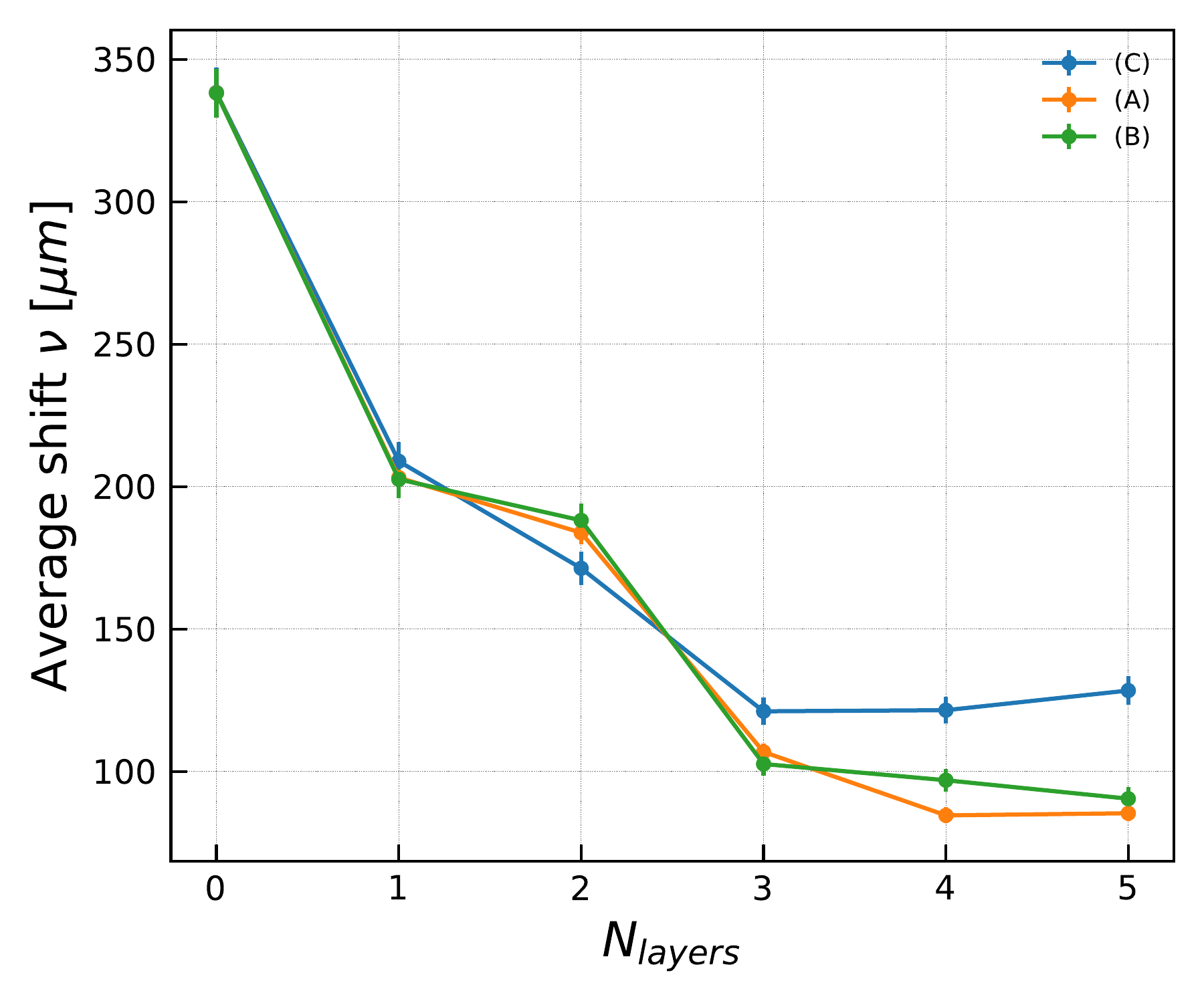}
    \includegraphics[width=0.32\linewidth]{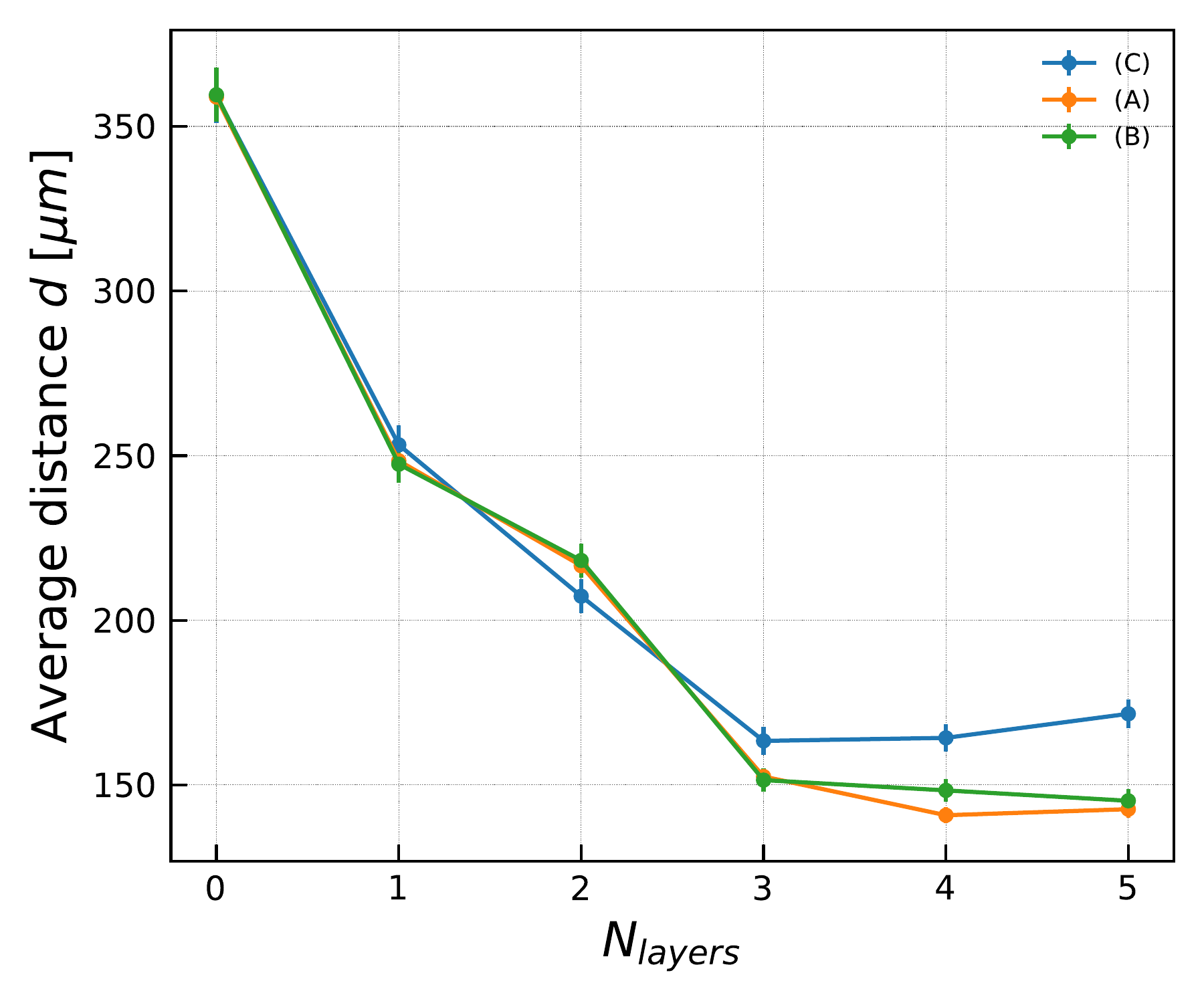}
    \caption{Mean resolution (left), shift (center) and distances (right) as function of the number of layers of the DNN for all splitting techniques and evaluated on the test sample. The mean is computed as the average over the scanned positions and the error bar represents the standard error of the mean.}
    \label{fig:averarge_performance_layers}
\end{figure}

\section{Conclusion}

We characterized the position reconstruction capabilities of an LG-SiPM comprising only 6 readout channels with a linear reconstruction technique and a DNN. We showed that the device has a granularity of about 0.69~mm nominally and that non-linearity can be recovered using a trained DNN granularity to about 0.20~mm when trained on a uniform random splitting of the data. In this case, the DNN reduces the number of distinguishable regions by a factor 12.1. The intrinsic granularity of LG-SiPMs, further enhanced by the DNN approach described in this work, is significantly finer than the pixelation achievable with current state-of-the-art scintillators used in scintillation cameras. This makes the choice of LG-SiPM suitable for applications requiring sub-millimetric resolution, especially with the implementation of Neural Network-based correction algorithms.

The present results are based on LED‑scan, we could study the improvements with real scintillation light used for radiation detection. The implications of scintillator size and optical spread, as well as the effect of microcell occupancy and potential saturation may modify the absolute position resolution when using real scintillators. However this do not affect the validity of the comparative analysis between the nominal (linear) reconstruction and the DNN approaches when considering only the intrinsic LG-SiPM capabilities. Future work will include measurements with scintillator‑coupled configurations to experimentally confirm the robustness of the DNN for radiation detection applications such as gamma-ray imaging. However it might be challenging, in a scintillator‑coupled system, to control the exact position and direction of incoming gamma‑rays due to the stochastic nature of radioactive decay, which limits the precision of beam‑based validation. As a consequence, precise ground‑truth position labels cannot be obtained experimentally, making LED‑based scans and Monte Carlo simulations he only practical way to perform supervised studies of reconstruction accuracy. This motivates the light‑only calibration strategy adopted in this work.

\section*{Acknowledgment}

The authors wish to acknowledge the University of Geneva and the Geneva University Hospitals for the support of the project administration and the use of their infrastructure. We also thank the Fondazione Bruno Kessler for providing the prototype LG-SiPM and for the support to test device capabilities

This project has received funding from the ATTRACT project funded by the EC under Grant Agreement 777222.

\bibliography{posics}

\end{document}